%Paper: hep-th/9510101
%From: Edward Witten <witten@sns.ias.edu>
%Date: Mon, 16 Oct 1995 16:42:59 EDT
%Date (revised): Mon, 23 Oct 1995 09:33:45 EDT

\input harvmac
\newcount\figno
\figno=0
\def\fig#1#2#3{
\par\begingroup\parindent=0pt\leftskip=1cm\rightskip=1cm\parindent=0pt
\baselineskip=11pt
\global\advance\figno by 1
\midinsert
\epsfxsize=#3
\centerline{\epsfbox{#2}}
\vskip 12pt
{\bf Fig. \the\figno:} #1\par
\endinsert\endgroup\par
}
\def\figlabel#1{\xdef#1{\the\figno}}
\def\encadremath#1{\vbox{\hrule\hbox{\vrule\kern8pt\vbox{\kern8pt
\hbox{$\displaystyle #1$}\kern8pt}
\kern8pt\vrule}\hrule}}

\overfullrule=0pt

%macros
%
\def\tilde{\widetilde}
\def\bar{\overline}

\font\zfont = cmss10 %scaled \magstep1

\def\bigone{\hbox{1\kern -.23em {\rm l}}}
\def\ZZ{\hbox{\zfont Z\kern-.4emZ}}

\Title{hep-th/9510101, IASSNS-HEP-95-78}
{\vbox{\centerline{SUPERSYMMETRIC YANG-MILLS THEORY}
\bigskip\centerline{AND INTEGRABLE
SYSTEMS}}}
\smallskip
\centerline{Ron Donagi\foot{Research supported in part by
NSF grant DMS95-03249
and Unite mixte de service de l'institut H. Poincare,
CNRS - Universite Pierre et Marie Curie, Paris.}}
\smallskip
\centerline{\it Department of Mathematics, University of Pennsylvania}
\centerline{\it Philadelphia, PA 19104-6395, USA}
\bigskip
\centerline{Edward Witten\foot{Research supported in part by
NSF grant PHY92-453.}}
\smallskip
\centerline{\it School of Natural Sciences, Institute for Advanced Study}
\centerline{\it Olden Lane, Princeton, NJ 08540, USA}\bigskip

\medskip

\noindent
%write abstract here
The Coulomb branch of $N=2$ supersymmetric gauge theories
in four dimensions is described in general by an integrable
Hamiltonian system in the holomorphic sense.  A natural construction
of such systems comes from two-dimensional gauge theory and
spectral curves.  Starting from this point of view, we propose
an integrable system relevant to the $N=2$ $SU(n)$ gauge theory with
a hypermultiplet in the adjoint representation, and offer much
evidence that it is correct.  The model has an $SL(2,{\bf Z})$
$S$-duality group (with the central element $-1$ of $SL(2,{\bf Z})$ acting
as charge conjugation); $SL(2,{\bf Z})$ permutes the Higgs,
confining, and oblique confining phases in the expected fashion.
We also study more exotic phases.
\Date{October, 1995}
%text of paper

\newsec{Introduction}

One of the basic objects of study in supersymmetric quantum
field theories in four dimensions is the moduli space of vacua.
For example, for $N=2$ supersymmetric Yang-Mills theories, there is
a Coulomb      branch in the moduli space consisting of vacua in
which the gauge      group $G$ is broken down to a maximal torus.
The Coulomb branch is of complex dimension equal to the rank $r$ of $G$.

The secret of the Coulomb branch \ref\paperi{N. Seiberg and E. Witten,
``Electric-Magnetic Duality, Monopole
Condensation, And Confinement In $N=2$ Supersymmetric
Yang-Mills Theory,'' Nucl. Phys. {\bf B426} (1994) 19.}
is that it parametrizes a family
of $r$-dimensional abelian varieties which
 controls the physics on this branch.  For instance,
masses of the stable massive particles are given by periods of a
certain differential form, and the most interesting physical
phenomena are determined by singularities of the abelian variety.
 An abelian variety
is a complex torus with a ``polarization'' which in the present
context comes from the symplectic pairing of electric and magnetic
charge.

\nref\paperii{N.
Seiberg and E. Witten, ``Monopoles, Duality, And
Chiral Symmetry Breaking In $N=2$ Supersymmetric QCD,'' Nucl. Phys.
{\bf B426} (1994) 484.}
For $G=SU(2)$, the rank $r$ is one and one is dealing with a family
of genus one Riemann surfaces depending on one complex parameter.
This is a simple enough situation that it has been possible
to deduce the structure rather directly
using some qualitative knowledge of the physics; this was done
in \refs{\paperi,\paperii} for the various $SU(2)$ theories
with zero or negative beta function.
\nref\argyres{P. Argyres and A. Faraggi, ``The Vacuum Structure And
Spectrum of $N=2$ Supersymmetric $SU(N)$ Gauge Theory,'' Phys. Rev.
Lett. {\bf 74} (1995) 3931.}
\nref\yankielowicz{A. Klemm, W. Leche, S. Theisen, and
S. Yankielowicz, ``Simple Singularities And $N=2$ Supersymmetric
Yang-Mills Theory,'' Phys. Lett. {\bf B344} (1995) 169.}
\nref\danielsson{U. H. Danielsson and B. Sundborg, ``The Moduli
Space And Monodromies Of $N=2$ Supersymmetric $SO(2r+1)$ Yang-Mills
Theory,'' hepth-9504102.}
\nref\landsteiner{A. Brandhuber and K. Landsteiner, ``On The
Monodromies Of $N=2$ Supersymmetric Yang-Mills Theory With
Gauge Group $SO(2n)$,       hepth-9507008.}
\nref\hanany{A. Hanany and Y. Oz, ``On The Quantum Moduli Space
Of $N=2$ Supersymmetric $SU(N_c)$ Gauge Theories,'' hepth-9505075.}
\nref\plesser{P. Argyres, R. Plesser, and A. Shapere, ``The Coulomb
hase Of $N=2$ Supersymmetric QCD,'' hepth-9505100.}
\nref\argsh{P. C. Argyres and A. D. Shapere,
``The Vacuum Structure Of $N=2$ Super-QCD With Classical Gauge
Groups,'' hep-th/9509175.}
\nref\hhanany{A. Hanany, ``On The Quantum Moduli Space Of Vacua
of $N=2$ Supersymmetric Gauge Theories,'' hep-th/9509176.}

A similar approach for $r>1$ would be extremely cumbersome,
and one naturally looks for a short-cut.  To date  solutions of
models with $r>1$ (with an exception mentioned at the end of
this introduction) have been based on assuming that one is looking
for the Jacobian (or in some cases, a Prym, the part of the
Jacobian odd under a ${\bf Z}_2$ symmetry) of a family of
hyperelliptic curves.  With some further qualitative assumptions and
physics input, it is then possible to determine the desired family.
This has been carried out very effectively for $SU(N)$
\refs{\argyres,\yankielowicz}, $SO(2N+1)$ \danielsson, and
$SO(2N)$ \landsteiner\ , all without matter, and  also
for models with matter fields in the fundamental representation
\refs{\hanany - \hhanany}.

In general, it is not clear which models can be described by   such a
hyperelliptic family, or indeed
by any family of Riemann surfaces (as opposed
to more general abelian varieties).  It is therefore natural to look
for other approaches.  In this paper, we start with the fact that
the total space of the sought-for family of abelian varieties is
a complex integrable system -  this assertion
will be clarified in section two.
Moreover, there is a natural gauge theory construction
\nref\hitchin{N. Hitchin, Stable Bundles and Integrable Systems,
 Duke Math. J. {\bf 54} (1987) 91-114.}
\nref\mark { E. Markman, Spectral Curves and Integrable Systems,
Comp. Math. {\bf 93} (1994) 255-290.}
\nref\DM {R. Donagi and E. Markman,  ``Spectral curves,
algebraically completely integrable Hamiltonian systems, and
moduli of bundles,'' 1993 CIME lecture notes, alg-geom-9507017,
 to appear in LNM. }
\refs{\hitchin - \DM}
of  integrable systems of just the right type.
So we have sought to match some of these
simple integrable systems  with $N=2$ models.

We have found one match, which will be the subject of this paper.
It is the $SU(n)$ theory with a matter hypermultiplet in the adjoint
representation; one can also think of this as the $N=4$ $SU(n)$
theory with a bare mass term  for some fields breaking
$N=4$ explicitly to $N=2$.

\def\Z{{\bf Z}}
$N=4$ super Yang-Mills theory was the original example that emerged
as a natural case for
 Montonen-Olive electric-magnetic
duality \ref\montonen{C. Montonen and D. Olive,
``Magnetic Monopoles As Gauge Particles?''
 Phys. Lett. {\bf 72B} (1977) 177. }.
This theory has zero beta function, so microscopically it
has a well-defined gauge coupling constant $e$.  There is no
anomalous $U(1)$ global symmetry, so the physics also depends
on a   vacuum angle $\theta$.
Montonen-Olive duality, originally formulated as an inversion of
the coupling constant, is extended when the theta angle is included
to an $SL(2,\Z)$ symmetry acting on
\eqn\hunj{\tau={\theta\over 2\pi}+{4\pi i\over e^2}.}
This generalization of the Montonen-Olive conjecture is called
$S$-duality.
The value of $\tau$ modulo the action of $SL(2,{\bf Z})$ precisely
determines the isomorphism class of an elliptic curve (genus one
Riemann surface) $E$.  $E$ can be described very explicitly
by an equation
\eqn\guggo{y^2=(x-e_1)(x-e_2)(x-e_3).}
We will build $S$-duality into our proposed solution by including
$E$ as part of the structure.   In addition to $x$ and $y$, we will
have a third variable $t$ and another equation
\eqn\uggo{F(t,x,y)=0}
with $F$ a rather special polynomial of degree $n$.  The equations
\guggo\ and \uggo\ for variables $x,y,$ and $t$ describe a complex
Riemann surface $C$ of genus $n$.  The solution of the model
is determined by the Jacobian of $C$ if the gauge group is $U(n)$;
for $SU(n)$ one considers the Jacobian modulo the part that comes
from $E$.

The precise polynomials
$F$ that appear here
would probably be rather hard to guess without the motivation
from the integrable system; thus, our approach differs from previous
investigations in that it is based on guessing a simple integrable
system which then determines the more complicated equations
of a Riemann surface,  rather than on guessing simple equations.
We hope that our approach will be useful for other examples, but
some ingredients are clearly still missing to give  a general
recipe for solving any $N=2$ model with any gauge group and
hypermultiplet representation.

\nref\vafa{C. Vafa and E. Witten, ``A Strong Coupling Test Of $S$
Duality,'' Nucl. Phys. {\bf B431} (1994) 3.}
As for the results that come from our solution of the model,
the most important point is certainly that $S$-duality is valid
for all $n$; previous computations focussed on $n=2$.
There is an interesting detail in how $S$-duality  is realized:
 the $S$-duality group turns
out to be $SL(2,\Z)$ rather than $PSL(2,\Z)$; the central
element $-1$ of $SL(2,\Z)$ acts by charge conjugation.

Also, the model has the beautiful property that (under a small
perturbation to $N=1$, as in \paperi) each and every
phase allowed by `t Hooft's classification of the massive
phases of $SU(n)$ gauge theory \ref\thooft{G. 't Hooft,  ``On The
Phase Transition Towards Permanent Quark Confinement,''
Nucl. Phys. {\bf B138} (1978) 1, ``A Property Of Electric
And Magnetic Flux In Nonabelian Gauge Theory,'' Nucl. Phys.
{\bf B153} (1979) 141,
``Topology Of The Gauge Condition And New Confinement Phases In
Nonabelian Gauge Theories,''
 Nucl. Phys. {\bf B190}  (1981) 455.}
occurs precisely once.  For general $n$,  such
phases are classified by subgroups of $\Z_n\times \Z_n$ of order
$n$, and the number of possible phases is the sum of the positive
divisors of $n$.  It emerges from
our solution that $SL(2,\Z)$ acts on the phases according to
its natural action on $\Z_n\times \Z_n$, as one would expect
by combining the ideas of `t Hooft with $S$-duality; for $n=2$
this has been seen in \refs{\paperii,\vafa}.  ($SL(2,{\bf Z})$ action
on analogous ${\bf Z}_n$ phases was first seen in a lattice
model \ref\cardy{J. L. Cardy and E. Rabinovici,
``Phase Structure of $Z(p)$ Models In The Presence Of
A Theta Parameter,'' Nucl. Phys. {\bf B205} (1982) 1.}

  Another interesting feature of the model is that it
exhibits singularities of a more general type than seen in previous
soluble models.  The simplest singularities of Riemann surfaces
are the nodes or ordinary double points, which, as explained in
\paperi,  lead to monopole
condensation and confinement.
In some ways, the next simplest singularity is the           cusp
$y^2=x^3$ which was argued  by Argyres and Douglas \ref\douglas{P. Argyres
and M. Douglas, ``New Phenomena In $SU(3)$  Supersymmetric Gauge
Theory,'' hepth-9503163.} to lead to a novel kind of superconformal
critical point in four dimensions.  These cusps show
up also in our model, and for gauge group $SU(3)$ we determine
how they are transformed by $SL(2,{\bf Z})$.
%Since the model we are studying flows, in the case of $SU(3)$,
%to the one studied in \douglas,
%it must generate at least the cusps seen there.
%What  we find is that within the 2-parameter family of $SU(3)$
%spectral curves there is a 1-parameter family of nodal curves,
%parametrized by points of an appropriate  discriminant curve.
%This discriminant, in turn, has:
%4 nodes, corresponding to the massive phases discussed above;
%8 cusps, each corresponding to a cuspidal spectral curve ;
%and one higher singularity, a tacnode, located at infinity.

Other singularities of curves
will lead to more general critical points.  For instance,
the singularity  $y^2=x^n$, which can readily arise in a family
of hyperelliptic curves, was also briefly discussed in \douglas.  Because
the Riemann surface $C$ in our construction is not hyperelliptic, still
more general singularities can arise.  It is very
plausible that  for sufficiently
large $n$ we can obtain an arbitrary singularity $F(x,y)=0$ of a plane
curve.

This paper is organized as follows.  In section two of the paper,
we explain why integrable systems are relevant and
describe the particular system that we believe controls the solution
of the $SU(N)$ gauge theory with a matter field in the
adjoint representation.
The candidate for the solution of the model that arises naturally
in the integrable system framework is put in an explicit
form  at the end of section two and then studied in section three.
The paper is written in such a way that interested readers can consult
section two, while those who are only interested in the solution of
the model can jump to section three, much  of which can be read
independently of section two.
In section four, we discuss a few points of physics that are needed
in interpreting some of the results of section three.

Perhaps the main potential interest of our construction is the
role of an auxiliary two-dimensional classical gauge theory that
is used in describing the solution of the four-dimensional
quantum gauge theory.  This may prove to have something to do with
an eventual better  understanding and explanation of duality.
Both the four-dimensional quantum gauge theory that we are solving
and the two-dimensional      auxiliary classical system have
fields known as Higgs fields that enter the formalism in rather
similar ways, though the reason for the close analogy is mysterious.

When this paper was substantially complete, there appeared
a paper by Martinec and Warner
\ref\mw{E. Martinec and N. Warner, ``Integrable Systems
and Supersymmetric Gauge Theory,'' hep-th/9509161.}
who used a certain class of
integrable models to describe pure $N=2$ gauge theory (without
hypermultiplets) with any simple gauge group.  We hope that construction
and the one presented here will prove to be special cases of
a more general story.

\newsec{Curves And Integrable Systems}

\subsec{Why Integrability?}

We first wish to explain what we mean in saying that the solution
of any of these $N=2$ models involves an integrable system in
the complex sense.

In \refs{\paperi,\paperii},  $SU(2)$ gauge theories were
solved in terms of a family
of elliptic curves $X\to U$, where $U$ is the complex $u$ plane
($u$ is the gauge-invariant order parameter $u=\Tr \,\phi^2$) and
the fibers of the map $X\to U$ are (except at finitely many singularities)
Riemann surfaces of genus one.  We write the fiber corresponding
to $u$ as $X_u$.  Part of the solution of the model involves also
giving a meromorphic differential one-form $\lambda$ on $X_u$, varying
holomorphically with $u$, such that the masses of the stable particles
are integer linear combinations of the fundamental periods $\vec a=
(a_D,a)$ of
$\lambda$.  In general, $\lambda$ has poles (the theory of which
can be quite elaborate; see section 17 of \paperii).  In addition,
$\lambda$ is not uniquely determined as a one-form;
 a transformation $\lambda\to\lambda+d\alpha$,
with $\alpha$  a meromorphic function, does not affect the periods.

There is no natural choice of $\lambda$.   Rather,
the ``gauge-invariant'' object, free of these ambiguities, is the
two-form $\omega=d\lambda$, which is moreover holomorphic.  Holomorphy
of $\omega$ is needed, as explained in section 6 of \paperi, to
prove positivity of the metric on $U$.  Though $\omega$ does
not quite determine the periods $\vec a$,  it determines their
derivative,
\eqn\unnd{d\vec a =\int_{\vec \gamma} \omega.}
On the left $d$ is the exterior derivative on  $U$, and
on the right $\vec\gamma$ is a set of fundamental one-cycles on the fiber.
\unnd\ is just a more abstract way to write the fact that as
$\vec a=\int_{\vec \gamma}\lambda$, $d\vec a/du=\int_{\vec \gamma}
d\lambda/du$.  As for the metric on $U$, it is defined as follows.
One starts with the $(2,2)$ form $\omega\wedge \bar\omega$ on
$X$. By ``integrating over the fiber'' of $X\to U$, one gets
a $(1,1)$ form on $U$, which is the Kahler form of the Kahler metric
on $U$.  This means concretely that if locally $\omega=\alpha\wedge du$,
with $\alpha$ a holomorphic
one-form on $X_u$, then the Kahler form on $U$
is $d\bar u \wedge du f(u)$ with $f=\int_{X_u}\alpha\wedge\bar \alpha$.
This is manifestly positive as long as $\alpha\not= 0$, that
is, as long as $\omega\not= 0$.

Now let us consider the generalization of this to the case of
a gauge group $G$ of rank $r>1$.  As explained in \paperi, section
3, the base $U$ is now a complex manifold of dimension $r$
(a copy of ${\bf C}^r$ parametrized by the gauge invariant order
parameters) and the physics is described by a family of $r$-dimensional
complex tori $X_u$, which are fibers of a map $X\to U$, with $X$
a complex manifold of dimension $2r$.  As described in \paperi,
the physics is then determined from a meromorphic one-form $\lambda$
whose restriction to the fibers is closed.  But
$\lambda$ has ambiguities and complicated
poles as just explained;
 the natural ``gauge-invariant'' object
is  $\omega=d\lambda$ which is a {\it closed holomorphic two-form}.
Concretely
\eqn\junn{\omega=\sum_idu^i {d\lambda\over du^i}}
where $u^i$ -- the gauge-invariant order parameters -- are coordinates
on $U$.  Because the restriction of $\lambda$ to the fibers of $X\to U$
is closed, $\omega$ is a sum of terms
each proportional to at least one one-form $du^i$ coming from the base;
in that sense, {\it the restriction of $\omega$
to the fibers of $X\to U$ is zero}.
The particle masses -- or rather their derivatives with respect
to $u^i$ -- are still given by \unnd.

To define the metric,
we need one further subtlety.  If one ignores the complex structure,
$X_u$ is a real torus of dimension $2r$ whose first homology is the
$2r$-dimensional space of magnetic and electric charges in this
theory (whose low energy gauge group is $U(1)^r$).
Given two particles with magnetic and electric charge
vectors $(\vec g_1,\vec e_1)$ and $(\vec g_2,\vec e_2)$, respectively,
there is a natural symplectic pairing $\vec g_1\cdot \vec e_2-\vec e_1
\cdot \vec g_2$ which according to the Dirac quantization law
is integer-valued.  This pairing is equivalent to a two-form
$t$ on $X_u$ which has integral periods and moreover (because of
the constraints of $N=2 $ supersymmetry) is positive and
of type $(1,1)$ in
the complex structure on $X_u$.  Such an object
defines what is called
a  ``polarization'' of $X_u$.
\nref\heat{E. Witten, ``Quantum Background Independence in String
Theory,'' hepth-9306122,
 in {\it Salamfest 1993}.}
\nref\cubics{R. Donagi and E. Markman, ``Cubics, Calabi-Yau threefolds,
Integrable Systems and Mirrors,'' alg-geom-9408004, to appear in
  Proc. Hirzebruch Conf.,
Israel Math. Conf. Proc. (AMS).}
\foot{If all electric and magnetic charges allowed
by Dirac quantization actually appear in the system, the polarization
will be principal; but this is not so in general. If one relaxes
the positivity condition on $t$, one gets an indefinite polarization
such as  arises typically in intermediate Jacobians of
higher dimensional varieties, cf. \refs{\heat,\cubics}, where
the varieties
are Calabi-Yau  threefolds.}  Endowed with this polarization,
our complex torus becomes an Abelian variety (which is simply a complex
torus that can be described by algebraic equations).

The metric can now be defined as follows: starting with
the $(r+1,r+1)$-form $t^{r-1}\wedge \omega\wedge \bar\omega$ on
$X$, integrate over the fibers of $X\to  U$ to get a $(1,1)$ form
 on $U$ which will be the Kahler form of the Kahler metric on $U$.
The metric obtained this way will always be positive semi-definite
and will be positive definite {\it if and only if $\omega$ is
non-degenerate}, that is if and only if in any local  coordinate
system on $X$, the matrix $\omega_{IJ},\,\,I,J=1\dots 2r$
of components of $\omega$
is invertible.  In fact, more concretely in our case non-degeneracy
means that locally one can pick the coordinates to be $u^i,\,i=1\dots r$
and some ``conjugate'' variables $x_i,\,i=1\dots r$
along the fibers, with
$\omega=\sum_i dx_i\wedge du^i$.  (If the restriction of $\omega$
to the fibers were non-zero, there would be additional terms $dx_i\wedge
dx_j$.)
$\omega$ being non-degenerate is equivalent to the fact that
all the $du^i$ appear in this formula; if one of them were absent,
the metric on $U$ would vanish in the corresponding direction and
so would not be strictly positive.  Thus, $\omega$ is non-degenerate
at least where the physics is non-singular; for the $SU(2)$ models,
$\omega$ is non-degenerate even at singular fibers, but it is not
clear whether this is general.

At least away from singular fibers, the basic structure is therefore
a family of abelian varieties $X\to U$ endowed with a
non-degenerate, closed, holomorphic two-form $\omega$, which moreover
has vanishing restriction to the  fibers.  If we just require
$\omega$ to be non-degenerate, closed, and holomorphic,
then it defines a {\it complex symplectic structure} on $X$.
This enables one to define Poisson brackets of holomorphic functions
much  as one usually defines Poisson brackets of functions on
an ordinary symplectic manifold.
Thus, denoting as $\omega^{IJ}$ the inverse matrix of $\omega_{IJ}$,
the Poisson bracket of two local holomorphic functions $f$ and $g $ is
defined by the usual formula
\eqn\lskk{\{f,g\}=\sum_{I,J}\omega^{IJ}\partial_If\,\partial_Jg.}
As in the usual case, the fact that $d\omega=0$ implies
that this Poisson bracket obeys  the Jacobi identity.
We have not yet used the fact that the restriction of $\omega$ to
the fibers vanishes.  This means, as above, that $\omega$ can
be written locally as $\sum_idx_i\wedge du^i$, and therefore
that {\it the Poisson brackets $\{u^i,u^j\}$ vanish}, that is,
{\it the $u^i$ are a maximal set of commuting Hamiltonians}.
This is the basis for asserting that we are dealing with the complex,
or even algebraic, analogue of a completely integrable Hamiltonian
system. We call such an object an algebraically
completely integrable Hamiltonian system.

Conversely, given an algebraically completely integrable Hamiltonian
system $X$, one can reconstruct much of this structure.
Defining $U$ to be the space parametrized by the commuting
Hamiltonians, one has a map $X\to U$ (which forgets the other
variables). If this map is proper, one can prove that the fibers are
complex tori  (otherwise we could get products of tori and affine spaces.)
In any case, $\omega$ has vanishing restriction to the fibers.

\bigskip\noindent
{\it Dependence On Mass Terms}

So far, we have described the structure that appears in a single
$N=2$ supersymmetric gauge theory.  Whenever hypermultiplets
are present, one gets a family of such theories, depending on certain
complex mass parameters $m_\lambda$.
Then $X$ and $\omega$ vary holomorphically
with the $m_\lambda$, subject to a condition explained in section 17 of
\paperii; the cohomology class $[\omega]$
of the  two-form $\omega$ varies
linearly in the $m_\lambda$.  This condition makes sense because, though
the complex structure of $X$ varies with the $m_\lambda$, as a real
manifold (of dimension $4r$) $X$ is fixed, and in particular its
real cohomology, where $[\omega]$ takes values, is locally a fixed vector
space.  (Globally, we are dealing with a vector bundle with flat
connection. Its  base,  parametrizing values of the mass parameters
$m_\lambda$ for which the corresponding $X$ is non-singular, is an open
subset in a vector space, while its fibers are the cohomologies of  those
$X$.)

The following very natural way to get a family $X,\omega$ with
$[\omega]$ varying linearly with respect to some parameters
will be used in the sequel.  Suppose that one is         given
a complex symplectic manifold $Y,\omega$ admitting the action
of a complex Lie group $H$.  The action of $H$ is generated by
holomorphic vector fields $V_a,\,a=1\dots {\rm dim}\,H$.  Just
as in ordinary classical mechanics, one can now look for holomorphic
functions $h_a$ that generate the action of $V_a$ by Poisson brackets.
(In components, this means, as usual,
 that $\partial_Ih_a=\omega_{IJ}V^J_a$.)
As in mechanics, one can now ``reduce'' with respect to the $H$ action,
the prototype being the reduction of the two-body problem with
a central force to a radial problem.  One does this by considering
only orbits with
\eqn\pillo{h_a=\mu_a,}
with the $\mu_a$ being some complex constants, and dividing by
the subgroup $H'$ of $H$ that leaves fixed the $\mu_a$.
(For instance, in mechanics $H$ might be the rotation group in the
two-body problem, the $h_a$ would be the components of
angular momentum, \pillo\
would assert that the angular momentum has a fixed value that points
in, say, the $z$ direction, and $H'$ would be the group of rotations
around the $z$ axis.)  The space of solutions of \pillo\ divided
by $H'$ is a complex manifold $X$ which is endowed with
a complex symplectic form induced from $\omega$
(which we will again refer to as $\omega$).  The
basic fact we need about this situation is that if one lets
the $\mu_a$ vary, keeping $H'$ fixed (in the example, with $H'$
the group of rotations around the $z$ axis, $\mu_a$ would be
an arbitrary vector pointing in the $z$ direction), then
{\it the symplectic structure of $\omega$ varies linearly with
the $\mu_a$} \ref\duist{J. J. Duistermaat and G. J. Heckman,
``On The Variation In The Cohomology In The Symplectic Form
Of The Reduced Phase Space,'' Invent. Math. {\bf 69} (1982) 259.}.
  (For real symplectic manifolds, this result enters
the theory of linear sigma models in an essential
way \ref\witten{E. Witten, ``Phases
Of $N=2$ Models In Two Dimensions,'' Nucl. Phys. {\bf B403} (1993) 159.}.)
When $X$ and $Y$ are related as above, $X$ is said to be a symplectic
quotient of $Y$ by $H'$.

Therefore, this setup, with the $\mu_a$ playing the role of the bare
masses in an $N=2$ supersymmetric gauge theory, gives a natural
way to automatically obey the constraint that $[\omega]$ varies
linearly in the masses.  As was seen in section 17 of \paperii,
that constraint is very powerful (completely determining, for instance,
the solution of the $SU(2)$ theory with $N_f=4$) but very complicated
to implement directly.  A construction that obeys this constraint
{\it a priori} is therefore highly desireable.

\def\C{{\bf C}}

\subsec{Gauge Theory And Integrable Systems}

In what follows, we use a natural construction of algebraically integrable
systems as symplectic quotients. The construction of such a system,
associated to an arbitrary Riemann surface and reductive group, is
due to Hitchin \hitchin. In order to obtain interesting {\it families} of
symplectic quotients, we use a version of this construction due to
Markman \refs{\mark, \DM} , which uses a marked Riemann surface
with  specified  conjugacy classes at the marked points.
Actually, we will only describe the small part of this beautiful
story that we need, and thus will not attempt to describe
the hyper-Kahler structure associated with these systems,
or the parts of the story that are purely topological in nature,
independent of complex structures.\foot{Also, we will glide
over various subtleties associated with unstable and semi-stable
points in quotients, so various assertions are only true on
dense open sets -- good enough for our purposes.}

We start with a complex Riemann surface $\Sigma$ and compact
gauge group $G$.  We let $A$ be a connection on a $G$-bundle over
$\Sigma$ -- for simplicity we take it to be the trivial bundle.
And we let $\Phi$ be a one-form with values in the adjoint representation
of $G$.  Let $Y$ be the space of such pairs.  We give $Y$ a complex
structure by saying that the $(0,1)$ part of $A$ and the $(1,0)$
part of $\Phi$ are holomorphic.  A complex symplectic structure is defined
on $Y$ by saying that the non-zero Poisson brackets of the (holomorphic)
components
of $A$ and $\Phi$ are
\eqn\unto{\{A_{\bar z}^a(x),\Phi_z^b(y)\}=\delta^{ab}\delta(x,y).}
Thus all components of  $\{A,A\}$ and $\{\Phi,\Phi\}$ vanish.

The  group $H_0$ of $G$-valued gauge transformations acts on $Y$ in the
usual fashion, preserving the complex structure.
Slightly less obvious is the fact that the complexification $H$ of $H_0$,
which is the group of $G_\C$-valued gauge transformations
($G_\C$ is the complexification of
 $G$) acts holomorphically on $Y$.
One simply takes $G_\C$ to act in the standard fashion on the holomorphic
fields, the $(0,1)$ part of $A$ and the $(1,0)$ part of $\Phi$,
while (since $A$ and $\Phi$ are real or hermitian by definition)
acting in the complex conjugate fashion on the complex conjugates
of these fields.  Let $H$ be the group of such $G_\C$-valued
gauge transformations.

Since $H$ acts holomorphically and preserving the complex structure,
one can ask what are the Hamiltonian functions that generate
the $H$ action under the Poisson brackets \unto.
 These turn out to be the objects
\eqn\nugo{ h=\bar D_A\phi}
where now $\bar D_A$ is the $\bar\partial $ operator determined by $A$
and $\phi$ is the $(1,0)$ part of $\Phi$.
\foot{To make this very explicit, for every adjoint-valued function
$\epsilon$ we have the infinitesimal    gauge transformation
$\delta A_i=-D_i\epsilon$, $\delta \Phi=[\epsilon,\Phi]$.  The
Hamiltonian $h(\epsilon)=\int_\Sigma \Tr \epsilon \bar D_A\phi$ generates
this transformation via the Poisson brackets defined above.}
Recall that the $\bar D_A$
operator gives a holomorphic structure to the trivial $G_\C$ bundle
over $\Sigma$, endowing it with a complex structure.
The symplectic quotient $X$ is thus the space of solutions
of $h=0$, that is
\eqn\nugog{\bar D_A\phi =0,}
divided by $H$.

Let us describe $X$ explicitly.  First of all we forget about $\phi$.
The $G_\C$ gauge transformations act on the $\bar D_A$ operator
by $\bar D_A\to g\bar D_A g^{-1}$, which is the usual equivalence
relation on $\bar\partial $ operators.  So the space of $A$'s
modulo the action of $H$ is the moduli space ${\cal M}$ of
holomorphic $G_\C$ bundles on $\Sigma$.  The cotangent space to
${\cal M}$ is the space of holomorphic one-forms valued in the
adjoint representation, that is, the space of $\phi$'s obeying
\nugog.  So the symplectic quotient
$X$ of  $Y$ by $H$  -- which should be  the space of solutions of
\nugog\ divided by $H$ -- is just the cotangent bundle $T^*{\cal M}$
of ${\cal M}$.\foot{In keeping with a previous footnote,
this is actually only a valid description of the symplectic quotient
on a dense open set.}  The symplectic structure of $X$  (obtained
by restricting \unto\ to modes tangent to solutions of \nugog)
 is just its
natural structure as a cotangent bundle.

Now (following Hitchin)
let us exhibit $X$ as a completely integrable system in the
complex sense.  To do this, if $X$ has complex dimension $2n$,
we must exhibit $n$ independent
holomorphic functions that Poisson-commute,
that is, $n$ commuting Hamiltonians.
Suppose first that $G=SU(2)$, and let
$\Sigma$ be of genus $g>1$.  Then $\dim_\C(X)=
6g-6$, so we need $3g-3$ commuting Hamiltonians.  To find them,
we simply observe that  $\Tr\,\phi^2$
is a holomorphic function on $Y$ (since it is a function only
of the $(1,0)$ part of $\Phi$) which moreover is gauge-invariant,
 so its restriction to $h=0$
descends to a holomorphic function on $X$.  Moreover, $\Tr\,\phi^2$
is a quadratic differential on $\Sigma$ which is holomorphic when
$h=0$,  Holomorphy of $\Tr\,\phi^2$ means that it can be paired
with $H^1(\Sigma,T)$ ($T$ is the holomorphic tangent bundle of $\Sigma$)
by integration. For $\alpha\in H^1(\Sigma,T)$, the formula
\eqn\udnn{v(\alpha)=\int_\Sigma \alpha\wedge \Tr \,\phi^2}
defines a holomorphic function on $X$.  As these functions
are constructed from $\phi$ only, they Poisson-commute, in view
of the structure of the Poisson brackets.  Since $H^1(\Sigma,T)$ has
dimension $3g-3$, we get the desired $3g-3$-dimensional space
of holomorphic Poisson-commuting functions.

For more general gauge groups, one repeats the above construction,
using all the independent gauge-invariant polynomials in  $\phi$
and not only the quadratic function $\Tr\,\phi^2$.  For instance,
for $SU(n)$, one uses the independent invariants $\Tr\,\phi^k$
for $k=2,3,\dots,n$.  Since $\Tr\,\phi^k$ is a holomorphic
$k$-differential on $\Sigma$, and the space of such holomorphic
$k$-differentials has dimension $(2k-1)(g-1)$ for $k>1$, one
gets  the correct number
$(3g-3)+(5g-5)+\dots +(2n-1)(g-1)=(n^2-1)(g-1)$ of  commuting
Hamiltonians.

\bigskip\noindent
{\it Reduction With Respect To A Subgroup}

One can generalize this and take the symplectic quotient with respect
to a finite-codimension subgroup $H'$ of $H$.  To do so, following
\refs{\mark}, we select
an arbitrary finite set of points $P_{(i)}$, $i=1,\dots , d$
on $\Sigma$.  At each $P_{(i)}$, pick an element $\mu_{(i)}$ of the
Lie algebra of $G$.  Instead of setting $h=0$, set $h$ to a sum
of delta functions supported at the $P_{(i)}$.  In view of the
definition of $h$, one does this by  imposing the equation
\eqn\unbn{\bar D_A\phi(x) = \sum_i\mu_{(i)}\delta(x,P_{(i)}).}
This condition means that $\phi$ is holomorphic away from the
$P_{(i)}$, and has simple poles at the $P_{(i)}$ with residue
equal to $\mu_{(i)}$.

Next we divide by $H'$, the subgroup of $H$ that commutes with
the $\mu_{(i)}$.  If we ignore $\phi$, then the quotient of the
space of connections by $H'$ is simply the moduli space
${\cal M}_{\vec \mu}$
of holomorphic $G_\C$ bundles on $\Sigma$ with the structure group
reduced at $P_{(i)}$ to the subgroup of $G_\C$ that commutes
with $\mu_{(i)}$.  Including $\phi$ and also imposing \unbn,
the symplectic quotient $X_{\vec\mu}$ of $Y$ by $H'$ is the space
of pairs consisting of a point in  ${\cal M}_{\vec\mu}$ and
a solution $\phi$ of \unbn.

Complete integrability is again established by looking
at the components of gauge-invariant pluri-differentials
such as $\Tr\,\phi^2$.
If we let the $\mu_{(i)}$ vary while keeping fixed $H'$,
then -- in keeping with a finite-dimensional phenomenon
explained above -- the cohomology class of the induced symplectic
structure on $X_{\vec\mu}$ varies linearly.

 So the general conditions are right
to describe an $N=2$ supersymmetric gauge theory in four dimensions
by such an integrable system, with the components of $\mu$ being
linear functions of the bare masses.

For readers interested in the classical geometry in more depth,
it may be helpful to compare the symplectic objects
$T^*{\cal M}_{\vec\mu}$ and $X_{\vec\mu}$. Over a general curve
$\Sigma$, let
$\pi:{\cal M}_{\vec\mu}\longrightarrow {\cal M}$
be the natural projection.  Its fibers are coadjoint orbits,
hence are naturally symplectic. The cotangent bundle
$T^*{\cal M}_{\vec\mu}$  has a corresponding subbundle
$\pi^*T^*{\cal M}$, which inherits a symplectic structure from
$T^*{\cal M}_{\vec\mu}$ and the coadjoint fibers. Our object
$X_{\vec\mu}$ looks locally like $\pi^*T^*{\cal M}$, in fact it is
an affine bundle over ${\cal M}_{\vec\mu}$ modelled on
the vector subbundle $\pi^*T^*{\cal M}$ of $T^*{\cal M}_{\vec\mu}$.
On the other hand, if $\Sigma$  is such that the general vector bundle
on it has a non-trivial group  $A$ of automorphisms, then the fibers of
$\pi$ are coadjoint orbits modulo $A$, and   $X_{\vec\mu}$ is now
modelled on the subquotient  $\pi^*T^*{\cal M}/A$ of
$T^*{\cal M}_{\vec\mu}$, which is still symplectic. (In the case of
interest to us, $\Sigma$ has genus 1,
so there is always a large $A$.)

\subsec{An Example}

As promised in the introduction, we will attempt to describe
in this way the $N=4$ theory with gauge group $SU(n)$
perturbed by a bare mass that reduces the symmetry to $N=2$.
First, though, we consider the $N=4$ theory without the mass
perturbation.

We need to pick a Riemann surface; what will it be?  Because the anomalies
vanish, this theory has a microscopic $\tau$ parameter
\eqn\jukko{\tau={\theta\over 2\pi}+{4\pi i\over e^2}.}
In seeking the solution, we will assume that the theory has
complete $SL(2,{\bf Z})$ invariance. The orbit
of $\tau$ modulo $SL(2,{\bf Z})$ precisely determines the isomorphism
class of a genus one curve $E$.  This gives the Riemann surface
we need.  We will proceed by constructing
integrable systems via gauge theory on $E$.

We also need a gauge group for the gauge theory on $E$. We will
simply use the same gauge group $G$ that entered the four-dimensional
problem we are trying to solve.

Using $E$ (without marked points) and $G$ in the above construction
of an integrable system, we get a definite integrable system with
nothing that can be adjusted.  Does it reproduce the vacuum
structure of the $N=4$ theory?
\foot{When we say ``vacuum structure'' here,
we are interested in those vacua that go over to the $N=2$ Coulomb branch
when the mass perturbation is added; we will ignore the expectation
values of fields that are part of the $N=2$ hypermultiplet.}
 We claim that it does (for any
compact $G$, and
not just the case $G=SU(N)$ to which we restrict later).

To see this, first take $G=U(1)$.  The curve $E$ can be described
by an explicit equation
\eqn\bbuf{y^2=(x-e_1)(x-e_2)(x-e_3)}
in the $x-y$ plane; the $e_i$ are distinct complex numbers.
For $G=U(1)$ and $E$ of genus one,
the moduli space ${\cal M}$ of flat $G_\C$ bundles on $E$ is just
a copy of $E$.
$\phi$ is supposed to be a holomorphic differential on $E$ with
values in the adjoint representation; because $G$ is abelian,
$\phi$ is just an ordinary holomorphic differential on $E$.
Any such differential is $\phi = a\phi_0$ with $\phi_0$ a fixed
holomorphic differential on $E$ and $a\in \C$.
The integrable system $X$ is therefore $E\times \C$, with $\C$
being the $a$-plane.  (As a check, note that $X$ should be
$T^*({\cal M})$, but with ${\cal M}$ being the torus $E$, $T^*{\cal M}$
is the same as $E\times \C$.)  The symplectic form
on $X$ is the natural symplectic form on $T^*({\cal M})$:
\eqn\knsn{\omega={dx\over y}\wedge da}
where $dx/y$ is a holomorphic differential on $E$.  The Kahler form
on the $a$-plane, obtained by taking $\omega\wedge\bar\omega$ and
integrating over $E$, is
\eqn\inno{{\rm Im}(\tau)\,da\wedge d\bar a.}
This flat metric is the correct free field theory metric on the $a$ plane
in the $N=4$ $U(1)$ theory.

Now we want to generalize this for non-abelian compact $G$ of rank $r$.
What is special about $N=4$ is that the correct answer for any $G$ is
in fact very similar to \inno, except that one must divide by the
Weyl group.  This holds because for $N=4$, the metric on the moduli
space of vacua  is completely determined by the symmetries, and in
particular is given exactly by the tree level expression.
Recall that $N=2$ relates the gauge   field to a complex scalar
field $\phi$ in the adjoint representation.  The vacuum is determined
by the value of $\phi$ up to conjugation by $G_\C$; this
means that a generic $\phi$ can be diagonalized, and replaced by
an $r$-tuple of complex
fields $\vec a=(a_1,\dots,a_r)$
taking values in a maximal abelian subalgebra of
the Lie algebra, except that one must identify two $\vec a$'s that
differ by action of the Weyl group.
The Kahler form -- read off    from the          classical Lagrangian
because there are no quantum corrections
 -- is the obvious generalization of
\inno, namely
\eqn\jinno{{\rm Im}\,(\tau)\,\Tr \,\,d\vec a\wedge d\bar{\vec a}.}
Dividing by the Weyl group $W$ of $G$,
this is to be understood as a metric
not on $\C^r$ but on the moduli space $U=\C^r/W$.

Let us try to reproduce this answer via gauge theory on $E$.
Because $\pi_1(E)$ is abelian, the moduli space of semistable holomorphic
$G_\C$ bundles on $E$ would be unchanged if one replaces $G$ by a maximal
torus $T$, except that one has to divide by the Weyl group.
Since $T=U(1)^r$, the moduli space of holomorphic
$G_\C$ bundles is $E^r/W$, the product of $r$ copies of $E$ divided
by $W$.  The $s^{th}$ copy of $E$  is described by an equation
\eqn\osnno{y_{s}^2=\prod_{i=1}^3(x_s-e_i)}
in the $x_s,y_s$ plane.
To this we must adjoin a holomorphic differential $\phi$ with
values in the adjoint representation.  At a generic point in $E^r$,
holomorphy forces the part of $\phi$ that does not commute with $T$
to be zero and the remaining part to be constant, so we get a complex
field $\phi=\vec a=(a_1,\dots,a_r)$ with values in the abelian
subalgebra, just as in the four-dimensional discussion in the last
paragraph.
The integrable system is thus $X=(E^r\times \C^r)/W$, with  $\C^r$
being the product of the complex $a_s$-planes, $s=1,\dots,r$; away
from the fixed points of $W$, this is the same as $X=T^*(E^r/W)$.
The symplectic form on $X$ (obtained by restricting the microscopic
Poisson brackets to solutions of \nugog) is
\eqn\knson{\sum_{s=1}^r{dx_s\over y_s}\wedge da_s.}
Computing from this   the metric (by integrating $t^{r-1}\wedge \omega
\wedge\bar\omega$ over the fibers
\foot{An elliptic curve $E$ has
a natural polarization $t_0$ -- the cohomology class dual to a point.
The polarization $t$ of $E^r$ that comes from the two-dimensional
gauge theory construction
is the sum of the polarizations on the factors. }),
we arrive at the desired metric
\jinno.  Just as in the four-dimensional
field theory, we interpret this as
a metric on $U=\C^r/W$, not on $\C^r$, because $X$ is really
$(E^r\times \C^r)/W$, not $E^r\times \C^r$.

\bigskip\noindent
{\it The Mass Perturbation}

Now we want to consider the mass perturbation breaking $N=4$ to
$N=2$.  This perturbation depends on precisely one complex parameter
$m$, and so we need a generalization of the above to a family of
complex integrable systems with cohomology class varying linearly
with $m$.
To do so, as motivated in the last subsection, we will let $\phi$
have poles, with residue linear in $m$.  We must pick the allowed
position and conjugacy classes of the poles.

There are two strong constraints.  First, we want to maintain the
full $SL(2,{\bf Z})$ invariance of the above construction.  For this,
we can single out a single point $P$ at which a pole will be allowed,
but singling out more than one point would break part of the
$SL(2,{\bf Z})$ symmetry.  For the genus one curve
\eqn\yusnn{y^2=\prod_{i=1}^3(x-e_i),}
we will take $P$ to be the point at $x=y=\infty$.  A genus
one curve with one point selected is called an elliptic curve.

What remains is to identify the $\mu$, that is,
the conjugacy class of the pole in
$\phi$ at $P$.  Here we run into our second strong constraint.
The integrable system $X$ just introduced has the correct dimension --
we saw above how the components of $\phi$ naturally turn into the
vacuum parameters on the Coulomb branch.  For generic $\mu$,
the integrable system $X_{\mu}$ has a dimension bigger than that
of $X$.  For example, for $G=SU(n)$, the case we consider in the
rest of this paper, there is a unique conjugacy class of $\mu$ (up to
scaling) that leads to $X_\mu$ of the correct dimension.
This is the case, considered in
\ref\TV{A. Treibich,  J.-L. Verdier, Solitons elliptiques, The
Grothendieck Festschrift, Vol. III, 437-480, Birkhauser Boston, 1990.}
in the context of elliptic solitons, in which $\mu$ is a diagonalizable
matrix whose eigenvalues are (a constant multiple of)
 $1,1,\dots,1,-(n-1)$; that is, $n-1$ eigenvalues
equal $1$, and one equals $-(n-1)$.   We therefore have precisely
one candidate for the solution of the $N=4$ $SU(n)$ theory with the
mass term; the rest of this paper is devoted to giving evidence
that it is correct.

The reason that allowing a pole of just this type does not increase
the dimension of $X$ is the following.  Recall that $X_\mu$ is modelled
on a subquotient $\pi^*T^*{\cal M}/A$ of  $T^*{\cal M}_{\mu}$, where $A$
is the group of automorphisms of a general vector bundle
($=(\C^*)^{n-1}$, in our case), and
$\pi:{\cal M}_{\mu}\longrightarrow {\cal M}$ is the natural projection.
In our case, the dimension of $A$ equals the fiber dimension of $\pi$
(equals $n-1$), so there is no change in the dimension of $X$. The key
fact is that the dimension of the coadjoint orbit of $\mu$ is twice the
dimension of $A$.

\subsec{Spectral Curves}

The  discussion may sound abstract, and one might despair of being
able to calculate.  What makes this possible is one more ingredient
in Hitchin's story, beyond what we have so far explained; this is the
notion of a {\it spectral cover}.  In explaining this, we will
take $G=SU(n)$, so $G_\C=SL(n,\C)$.

A point in the integrable system $X$ is a holomorphic $G_\C$
bundle $V$  together with an adjoint-valued holomorphic one-form
$\phi$ on the Riemann surface $\Sigma$ of genus $g$.
Consider the $n$-sheeted cover of the genus-$g$ surface $\Sigma$
given by the equation
\eqn\snsnp{\det (t-\phi) = 0}
where $t$ takes values in the canonical line bundle of $\Sigma$.
This gives a Riemann surface $C$, of genus
$\widetilde{g} :=  (n^2-1)(g-1) + g $.
The claim is that everything can be described from the structure of
$C$.  This makes it possible to reduce
the story from a general discussion of families of abelian varieties
to very concrete issues about Riemann surfaces given by concrete
equations.

Note first of all that there is no problem in making concrete
the equation \snsnp.  This equation can be written out  very
explicitly
\eqn\tnns{t^n+t^{n-2}
W_2(\phi)+t^{n-3}W_3(\phi)+\dots +W_n(\phi)=0.}
The $W_k(\phi)$ are concrete gauge-invariant polynomials in $\phi$,
and are, in fact, holomorphic $k$-differentials on $\Sigma$
(possibly with poles of a specified kind).  If one expands
the $W_k$ in a basis of $k$-differentials, the coefficients that
arise are just the values of the commuting Hamiltonians of the
integrable system.

What remains is to understand the fibers of the
map $X\to U$ from the integrable system $X$ to the     space $U$ of
commuting Hamiltonians.
Over a generic point $ w\in \Sigma$, the equation \snsnp\ has
$n$ distinct solutions for $t$,
corresponding to one-dimensional eigenspaces
of $\phi$.  Let $v$ be a general point in $C$.  It sits over some
point $w\in \Sigma$, and corresponds to a one-dimensional eigenspace
of $\phi(w)$.  Call this  eigenspace $L_v$.  $L_v$ varies
holomorphically with $v$, as the fibers of a holomorphic line
bundle ${\cal L}\to C$.

{}From ${\cal L}$, the $SL(n,\C)$ bundle $V\to \Sigma$ can be
reconstructed as follows: if the distinct points lying over $w\in \Sigma$
are $v_i(w)$, then the fiber of $V$ over $w$ is
\eqn\mimmk{V_w=\oplus_{i=1}^n L_{v_i(w)}.}
This assertion is just the reconstruction of $V_w$ as the sum of the
eigenspaces of $\phi$.  If $V$ were simply      a $GL(n,\C)$ bundle
(as would be the case for gauge group $G=U(n)$) we would stop here:
every $V\to \Sigma$ gives a line bundle ${\cal L}\to C$, and conversely
by \mimmk,
so the fiber of the map $X\to U$ would be the Jacobian of $C$.

For gauge group $G=SU(n)$, we want
 $V$ to be an $SL(n,\C)$ bundle, which means that the  determinant
line bundle $\det(V)$ of $V$ should be trivial.
Given a line bundle ${\cal L}$ on $C$, we can define a holomorphic
line bundle $N({\cal L})$ on $\Sigma$ whose fiber
at  $w\in \Sigma$ is
\eqn\immko{N({\cal L})_w=\otimes_{i=1}^n L_{v_i(w)},}
If $V$ is as in \mimmk, then
$\det(V)=N({\cal L}) \otimes {K_{\Sigma}}^{\otimes n(n-1)/2}$, cf.
\ref\MSRI{R. Donagi, ``Spectral Covers,'' alg-geom-9505009,
to appear in the proceedings
of the MSRI special year in algebraic geometry (1992-93).};
hence the condition that $V$ is an $SL(n,\C)$ bundle is that
 $N({\cal L})$ should be the appropriate multiple of $K_{\Sigma}$.
So the fiber of the map $X \to U$ is, up to shifts,  the
kernel of the map ${\cal L}\to N({\cal L})$ from the Jacobian
of $C$ to that of $\Sigma$. In the case that $\Sigma$ is elliptic,
there will be no shift, as $K_{\Sigma}$ is trivial.

\bigskip\noindent
{\it Detailed Recipe       }

For the case we actually want to look at, $\Sigma$ is the genus
one curve defined by
\eqn\unsnon{y^2=\prod_{i=1}^3(x-e_i)}
for some complex numbers $e_i$, and $\phi$ is permitted to have
a simple pole at the point $P$ with
$x=y=\infty$.  The residue of $\phi$ is to be
a matrix with eigenvalues a constant times $1,1,\dots,1,-(n-1)$.
The constant is a multiple of the bare mass of the hypermultiplet.
Because the $N=4$ theory is scale-invariant, the actual value of
the bare mass (given that it is non-zero) and thus of the
constant in the residues does not matter.
\foot{To be more precise, the absolute value of $m$ can be removed
by scaling, and the phase by one of the $SU(4)$ global symmetries
of the $N=4$ theory.}

To work near $P$, we set $x=u^{-2}$ and $ y=vu^{-3}$, where $u$ is
a local parameter near $P$ and
$v$ is holomorphic at $u=0$.  We fix the normalization of the poles
in $\phi$ by requiring that near $u=0$, the polar part of $\phi$
is precisely $1/u$ times a matrix of eigenvalues $1,1,\dots,1,-(n-1)$.
The spectral cover $C$ of $E$ is given by an equation
$\det(t-\phi)=0$.  Setting $F(t,x,y)=\det(t-\phi)$, we
have
\eqn\snum{F(t,x,y)= t^n+B_2(x,y)t^{n-2}+\dots +B_n(x,y),}
where the $B_k$ are polynomials in $x$ and $y$ (so as to have no
singularities except at $P$) and
grow at most as $u^{-k}$ for  $u\to 0$.  Moreover, if we regard
the equation $F(t,x,y)=0$ as an equation for $t$ with $x$ and $y$
fixed, then
of the $n$ roots, $n-1$ have
$t$ growing as $1\cdot u^{-1}$ near $u=0$, and one root behaves as
$-(n-1)u^{-1}$.  This is because the roots of $\det(t-\phi)=0$ are
just the eigenvalues of $\phi$, and their growth for $u\to 0$ is
determined by the polar part of $\phi$.

The  condition on behavior of the roots
means that if we set $t'=t-u^{-1}$, then
only one solution for $t'$ has a pole at $u=0$.  That means that near
$u=0$
\eqn\ikob{F(t',u)=F_0(t',u)+{1\over u}F_1(t',u)}
with $F_0$ and $F_1$ holomorphic at $u=0$.
As we will see at the beginning of section three, this information,
together with the facts mentioned in the last paragraph, completely
determines $F$ -- and therefore the spectral curve  $C$ -- in terms
of the expected parameters.
For now, though, we interrupt the derivation  so that readers
who have chosen to omit the present section can rejoin us in section
three.

\newsec{Properties Of The Solution}

Our proposal for solving the four-dimensional $SU(n)$ gauge theory with
a massive hypermultiplet in the adjoint representation
(we will set the bare mass to one and not mention it explicitly)
involves a Riemann surface
$C$ described by writing two equations for three variables
$x,$ $y$, and $t$.  One is the equation
\eqn\ikn{y^2=(x-e_1)(x-e_2)(x-e_3)}
for a genus one Riemann surface $E$ whose $\tau$ parameter
should equal $\theta/2\pi +4\pi i/e^2$, with $e$ and $\theta$
the coupling constant and theta angle of the theory that we wish
to solve.

The second equation is
\eqn\unsn{F(t,x,y)=0 }
where $F$ will be described presently.  Together with \ikn, $F$ describes
a curve $C$ of genus $n$.  The physics is described by the part of the
Jacobian of $C$ that does not come from $E$.

The first condition on $F$ is that $F$
is a polynomial in $t,x$, and $y$ which -- if we consider
$x,y$, and $t$ to be of degree $2,3$, and $1$ respectively -- contains
only terms of degree $\leq n$.
We also may as well assume (since we are imposing also \ikn) that
$F$ contains only terms at most linear in $y$.
Moreover, in its $t$-dependence,
$F=t^n+O(t^{n-2})$.  These conditions would leave many free parameters
in $F$.  To fix them, we need a further condition on the behavior
of $F$ for $x,y\to\infty$.  Near infinity, write $x=u^{-2}$,
$y=u^{-3}v$, with
\eqn\sonn{v^2=\prod_{i=1}^3(1-e_iu^2).}
Thus, in particular, $v$ is holomorphic near $u=0$ and
has value $v=1$ there; $u$ is a good
local parameter near $x=y=\infty$.  The last condition
is that if $F$  is written in terms of $u$ and $t'=t-u^{-1}$,
then         $F(t',u)$ has only a first order pole at $u=0$, that
is,
\eqn\inns{F=F_0(t',u)+u^{-1} F_1(t',u)}
where $F_0$ and $F_1$ are holomorphic at $u=0$.  These conditions
together determine $F$ in terms of $n-1$ complex parameters; these
are the expected parameters on the Coulomb branch.

In fact, it turns out that if we further required that $F(t,0,0)=t^n$,
there would be  a unique polynomial $F$ obeying the constraints
of the previous paragraph.  There is one such polynomial for each $n$;
let us call it $P_n$.  Then the general $F$ allowed by the
conditions of the previous paragraph is
\eqn\hsnnn{F=P_n+A_2P_{n-2}+A_3P_{n-3}+\dots +A_nP_0.}
Here the $A_i$ are complex constants which we want to identify with
the order parameters on the Coulomb  branch.
That is, the $A_i $ are gauge invariant polynomials in the scalar
field $\phi$ that is related to the $SU(n)$ gauge field by $N=2$
supersymmetry. In the massless theory as analyzed in section two,
the $A_i$ are the $i^{th}$ elementary symmetric functions of $\phi$
(up to multiplicative constants that depend on how $\phi$ is normalized);
depending on the precise formalism,
there may be  mass-dependent corrections to this.
(In the formalism of section 2, there is another object defined
on the Riemann surface $E$,  curiously
also often  called a Higgs field, whose characteristic polynomial also
involves the $A_i$.  The relation between the two needs a better
explanation.)

Let us work out explicitly the first few $P_k$.  For $P_0$ and
$P_1$, it is impossible to have any terms at all involving $x$ and $y$,
since they have degree $\geq 2$, so
\eqn\sinni{\eqalign{P_0 & = 1 \cr
                    P_1 & = t .\cr}}
For $P_2$, the general polynomial obeying all conditions except
the behavior at $u=0$ is $t^2+\alpha x$ with $\alpha$ a constant.
Imposing the condition that $P_2(t',u)$ has only a simple pole
at $u=0$, we get
\eqn\inni{P_2=t^2-x.}
Similarly, starting with $P_3=t^3+\alpha xt +\beta y+\gamma x$,
we get
\eqn\pinni{P_3=t^3-3xt+2y.}
To go farther, it is useful to note that the conditions by which
the $P_n$ are determined imply that $dP_n/dt=nP_{n-1}$, so that
once $P_{n-1}$ is known, one knows $P_n$ modulo a polynomial
in $x$ and $y$ only.  With a little more work one gets
\eqn\inni{\eqalign{P_4 & = t^4-6xt^2+8yt-3x^2+4x\sum_ie_i\cr
                   P_5 & = t^5-10xt^3+20yt^2+(-15x^2+20x\sum_ie_i)t
              +4xy-8y\sum_ie_i}}
and so on.

Though these formulas will suffice for our applications,
the general form of the $P_n$'s can be described
 more systematically as follows. Write
$$P_n(t) = \sum_i\left (\matrix{
n\cr
i\cr
}\right )
f_{i,n}(u)t^{n-i}.$$
The condition $nP_{n-1}=P_n'$ gives $f_{i,n}=f_{i,n-1}=:f_i$,
independent of n. Our main condition is that, substituting $t=t'+1/u$,
$P_n$ should have pole order $\leq 1$ in $u$. In particular, setting
$t'=0$, i.e. $t=1/u$, we find that the following expression can have
at worst first order poles in $u$:
\eqn\jdjj{\eqalign{     \sum_n{{1 \over {n!}}P_n({1 \over u})s^n}= &
\sum_{n,i}{{1 \over {i!(n-i)!}} {f_i(u) \over {u^{n-i}}} s^n}=
\sum_{i,j}{{1 \over {i!j!}} {f_i(u) \over {u^j}} s^{i+j}} \cr & =
(\sum_i{{f_i(u) \over {i!}} s^i}) (\sum_j{ (s/u)^j \over j!})
=e^{s/u}F(u,s),}}
where $F(u,s):=\sum_i{{f_i(u) \over {i!}} s^i}$
is a holomorphic function $F: (E \setminus \infty) \times \C \to \C$
satisfying (and uniquely characterized by):
 (i) $F(su,s)$ is holomorphic in $s$ (i.e. $({\partial \over \partial
s})^i{F(u,0)}$
has pole order $\leq i$ in $u$);
  (ii) $e^{s/u}F(u,s)$ has pole order $\leq 1$ in $u$.
These conditions should translate into an explicit formula
for the generating function $F$ in terms of theta functions
and exponential terms on $E$.

\bigskip\noindent
{\it $S$-Duality}

It is now straightforward to exhibit the $S$-duality of the
formalism.  According to classical formulas
used in section 16 of \paperii, introduce the
theta functions
\eqn\thetaf{\eqalign{ \theta_1(\tau) & =\sum_{n\in {\bf Z}}q^{{1\over 2}
                      (n+1)^2} \cr
                      \theta_2(\tau) & = \sum_{n\in {\bf Z}}(-1)^n
                                          q^{{1\over 2}n^2} \cr
                      \theta_3(\tau) & =    \sum_{n\in {\bf Z}}
                                                 q^{{1\over 2}n^2},}}
with $q={e^{2\pi i\tau}}$, and set
\eqn\geta{\eqalign{e_3-e_2 & = \theta_1^4(\tau)  \cr
                   e_1-e_3 & = \theta_2^4(\tau) .  \cr
                    e_1-e_2 & = \theta_3^4(\tau)  \cr}}
With this choice, \ikn\ describes an elliptic curve whose
$\tau$ parameter is in fact $\tau$, and the coefficients in \ikn\
transform as modular forms of $SL(2,{\bf Z})$ (one
has $4\prod_i(x-e_i)=4x^3-g_2x-g_3$ where $g_2$ and $g_3$ are
Eisenstein series defining modular forms of weight four and six
for $SL(2,{\bf Z})$).  \ikn\ is therefore invariant  under the
action of $SL(2,{\bf Z})$ on $\tau$ if $x$ and $y$ are taken to
transform as modular forms of weight two and three, respectively.

The modular covariance extends to our second equation $F(t,x,y)=0$,
if we take $t$ to transform as a modular form of weight one, and
$A_k$ to transform as a modular form of weight $k$.  This result
about the modular weight of the $A_k$ extends the result in
\paperii\ for $n=2$, where $u=A_2$ was found to transform with
weight two.

Having $A_k$ transform with weight $k$ means that under
$\tau\to (a\tau+b)/(c\tau+d)$,
\eqn\aktrans{A_k\to {A_k\over (c\tau+d)^k}.}
In particular, under the element $-1$ of the center of $SL(2,{\bf Z})$,
one has
\eqn\jutrans{A_k\to (-1)^kA_k.}
This operation is usually called charge conjugation.  We have learned,
then, that the $S$-duality group is really $SL(2,{\bf Z})$ rather
than $PSL(2,{\bf Z})$, with the center acting by charge conjugation.
For $n=2$, the center acts trivially on the gauge-invariant
order parameters, though in a sense one still sees the charge
conjugation by the action of $SL(2,{\bf Z})$ on $y$.

\subsec{Flow To The Pure $N=2$ Theory}

In the rest of this section, we extract other key properties of
the solution and compare to what is known about the model independently.
The first issue that we will consider is the flow to the pure $N=2$
theory.  In other words, in a limit in which the bare mass goes
to infinity, and $\tau$ goes to infinity as the logarithm of the
mass, and one also performs suitable renormalizations of the $A_k$,
the $N=2$ theory with the matter hypermultiplet should reduce to
the pure $N=2$ theory with gauge group $SU(n)$,
whose structure is already known.  Let us verify this.   Because
of scale invariance, instead of taking the bare mass to infinity,
we can keep the bare mass at one and take the $A_k$ to ``zero'';
we put the word ``zero'' in quotes because some additive renormalization
may be involved.

We have found that the desired flow can be exhibited very simply
if one takes the equation for $E$ to be
\eqn\jurry{y^2=x(x-1)(x-\lambda),}
with some complex $\lambda$; any $E$ can be put in this form
by an affine transformation $x\to ax+b$ mapping $e_1,e_2,e_3$
to $0,1,\lambda$.
Note that the    affine transformation from \geta\ to \jurry\
leaves invariant the conditions (such as the fact that $F(t',u)$ has
only a simple pole at $u=0$) that characterized the allowed
$F$'s.  However, it does not leave invariant the further condition
$P_k(t,0,0)=t^k$ that was used to define the $P_k$'s.  Instead,
 $x\to ax+b$ will add to  $P_k$ a linear combination of $P_r$'s for
$r<k$, and therefore likewise will add to $A_k$  a linear combination of
 $A_r$ for $r<k$.  This generalizes the situation found in
section 16 of \paperii\ where the definition of $\Tr\,\phi^2$ needed
to exhibit $SL(2,{\bf Z})$ invariance differed by an additive
renormalization from the definition that was natural to exhibit
the flow to the pure $N=2$ theory.  (As $A_2=\Tr\,\phi^2$ and
$A_0=1$, an additive
renormalization of $\Tr\, \phi^2$ is a special case of a transformation
adding to $A_k$ a linear combination of $A_r$ for $r<k$.)  We will
denote the new $A_k$'s which arise if \hsnnn\ is used together
with \jurry\ as $A_k'$.

We can take the weak coupling limit to be $\lambda\to 0$.
In that limit, we take $x=\lambda \tilde x$, $y=\lambda \tilde y$,
where $\tilde x$ and $\tilde y$ are to have limits as $\lambda\to 0$.
The equation for $E$ thus reduces to
\eqn\murry{\tilde y^2= -(\tilde x -{1\over 2})^2 +{1\over 4}.}
We now have to look at the second equation
$F(t,x,y)=0$.  Though a general algebrogeometric argument exhibiting
the desired limit for all $n$ is possible, we will here show how
it works explicitly for small $n$.
Let us recall that the solution of the pure $N=2$ theory for $SU(n)$
can be described \refs{\argyres,\yankielowicz}  by the curve
\eqn\omigo{w^2=(z^n+b_2 z^{n-2}+b_3z^{n-3}+\dots +b_n)^2+1}
in the $w-z$ plane, with the $b_k$ being the order parameters.

For $n=2$, that is for $SU(2)$, our equation $F=0$ is
\eqn\two{t^2-x+A_2'=0.}
We set $t=\lambda^{1/2}\tilde t$, $A_2'=\lambda a_2$,
so that
this reduces to $ \tilde t^2-\tilde x + a_2=0$.  Solving
for $x$ and inserting in \murry, we get
\eqn\twotwo{\tilde y^2=-
\left(\tilde t^2-{1\over 2}+a_2\right)^2+{1\over 4}}
which with obvious substitutions (including a shift of $a_2$ by $1/2$)
is equivalent to \omigo\ for $n=2$.

For $n=3$, the equation $F=0$ reads
\eqn\three{t^3-3xt+2y+A_2't+A_3'=0.}
Now we take $t=\lambda^{1/3}\tilde t$, $A_k'=\lambda^{k/3}a_k$.
The limit of \three\ is then
\eqn\threethree{\tilde t^3 +a_2\tilde t+a_3 +2\tilde y=0.}
Solving for  $\tilde y$ and substituting in \murry,
the limit as  $\lambda\to 0$ is
\eqn\threeoth{(\tilde t^3+a_2\tilde t +a_3)^2=-4(\tilde x-{1\over 2})^2
+1}
which again is equivalent to \omigo\ with obvious substitutions.

The general pattern continues like this for larger $n$.  One
scales $t=\lambda^{1/n}\tilde t$, $A_k'=\lambda^{k/n}a_k$, and
one eliminates $\tilde x$ or $\tilde y$ depending
on whether $n$ is even or odd.  In the limit $\lambda\to 0$
one gets \omigo\ after obvious
substitutions.  Note that the relation between $\tau$ and $\lambda$
for small $\lambda$ is $ \lambda=q^{1/2}=e^{\pi i \tau}$, so our
scaling involves $A_k'=q^{k/2n}a_k$.  In fact, according to the
renormalization group, the mass scale $\Lambda$ of the pure $N=2$ theory
is related to the bare mass $m$ by $\Lambda^{2n}=qm^{2n}$ so
(as we have set $m=1$) our relation is $A_k'=\Lambda^k a_k$, as expected;
in other words, to flow to the pure $N=2$ theory, we must take $q\to 0$
while keeping the $A_k'$ (whose dimension is $k$)
fixed in units of $\Lambda^k$.

\subsec{Exact Description For $n=2$}

Another important check of our solution is to compare it to the
results already known for $SU(2)$, that is for $n=2$.

For $n=2$, the curve $C$ is described by the equations
\eqn\ilpo{\eqalign{y^2 & = (x-e_1)(x-e_2)(x-e_3) \cr
                                      0 & = t^2-x+A_2 .\cr}}
$C$ has a ${\bf Z}_2\times {\bf Z}_2 $ symmetry, with the first
${\bf Z}_2$ generated by $\alpha:y\to -y$ and the second
generated by $\beta:t\to -t$.

$C$ is a curve of genus two, so its Jacobian is two-dimensional.
We need to split off from the Jacobian of $C$ a one-dimensional
piece that will be used to describe the model.  The general recipe
was explained at the end of section two, but in the present
case a simplification is possible: we want the part of the Jacobian of $C$
that is invariant under the diagonal transformation $\alpha\beta$,
which changes the sign of both $y$ and $t$.
(This kills the period that comes from $E$ -- which is associated
with the differential form $dx/y$ that is odd in $y$ and even in $t$ --
and leaves the other period.)

This piece of the Jacobian
of $C$
is just the Jacobian of the curve  obtained by considering only
the $\alpha\beta$-invariant functions of $x,y,$ and $t$, subject
to the equations \ilpo.  For the basic $\alpha\beta$-invariant
functions, we can pick $x,z=t^2$, and $w=yt$.  After using the second
equation in \ilpo\ to eliminate $z$, the first becomes
\eqn\kilpo{w^2=(x-A_2)(x-e_1)(x-e_2)(x-e_3).}
According to our proposal for the $SU(n)$ theory, the $SU(2)$ theory
should be described by the Jacobian of this curve $D$.

Instead, in \paperii, the $SU(2)$ theory was described by
\eqn\vilpo{y^2=(x-e_1u-e_1^2)(x-e_2u-e_2^2)(x-e_3u-e_3^2).}
A small computation shows that the cross-ratio of the four points
$A_2,e_1,e_2,e_3$ agrees with that of the four points $\infty,
e_1u+e_1^2,e_2u+e_2^2,e_3u+e_3^2$ provided
$A_2=u+e_1+e_2+e_3$, so that \kilpo\ can be mapped to \vilpo\ by
an $SL(2,\C)$ transformation of $x$   given  that relation
between $A_2$ and $u$.  In the manifestly $S$-dual formalism
of \geta, $e_1+e_2+e_3=0$ and $A_2=u$; more generally the two
descriptions in \kilpo, \vilpo\ differ by an additive renormalization
of $u=\Tr\,\phi^2$.

\bigskip\noindent{\it Elimination Of $y$ For $n>2$}

For $n=2$, we were able to eliminate one variable (namely $x$) and
exhibit $C$ as a curve in the $y-t$ plane.  Is there any analog
of this for $n>2$?
For $n=3$, the equation $F(t,x,y)=0$ is explicitly
\eqn\forthree{t^3-3xt+2y+A_2t+A_3=0.}
One can solve this for $y$, and then substitute in the equation
for $E$.  One learns that the curve $C$ of the $SU(3)$ theory is
 the curve
\eqn\threecurve{{1\over 4}\left(t^3-3xt+A_2t+A_3\right)^2
-(x-e_1)(x-e_2)(x-e_3)=0,}
in the $x-t$ plane, a description that will be useful later.
Since $F$ is always linear in $y$, one can for $n>3$ always
solve $F=0$ to give $y$ as a rational function (not a polynomial
if $n>3$) in $x$ and $t$; substituting in the equation
$y^2=\prod_i(x-e_i)$  then exhibits $C$  as a rather complicated
curve in the $x-t$ plane.

\subsec{Singularities For Weak Coupling}

Another check comes by comparing to the singularity structure
of the theory in the weak coupling limit.  Of course, we have
already looked at the limit of weak coupling with the $A_k$ going
to ``zero''; now we will look at weak coupling with the $A_k$ fixed.

In general, a  singularity occurs for values of the $A_k$ at
which an extra massless particle appears.  In the classical limit,
the mass spectrum can be read off from the classical Lagrangian.
If the Higgs field $\phi$ has eigenvalues $a_1,\dots,a_n$
(with $\sum_ia_i=0$), then in the classical
limit, the masses of the vector supermultiplets
that are not massless generically equal
 $|a_i-a_j|$ with $i\not= j$, and the masses of the hypermultiplets
are (if the bare mass is 1) $|a_i-a_j-1|$.  A singularity will
arise from a zero of either
\eqn\imvo{D_V=\prod_{1\leq i<j\leq n}(a_i-a_j)^2}
or
\eqn\jimvo{D_H=\prod_{1\leq i,j\leq n}(a_i-a_j-1)
=(-1)^{n(n+1)/2}\prod_{1\leq i<j\leq n}\left((a_i-a_j)^2-1\right).}

We will study the weak coupling behavior in detail for $n=3$.
For $n=3$, if
\eqn\ppo{\eqalign{W_2=& a_1a_2+a_2a_3+a_3a_1\cr
                  W_3 & = a_1a_2a_3 \cr}}
are the elementary symmetric polynomials in the eigenvalues of
$\phi$, then
one has explicitly
\eqn\pimvo{\eqalign{D_V& =-4W_2^3-27W_3^2 \cr
                    D_H& =-(4W_2+1)
                    (W_2+1)^2-27W_3^2.\cr}}

On the other hand, if we denote the left hand side of \threecurve\
as $Q(x,t)$, for $n=3$ a singularity of $C$ arises precisely when
$Q=\partial Q/\partial x = \partial Q/\partial t = 0$.
The equation $\partial Q/\partial t = 0 $ gives\foot{There is
another branch $t^3-3xt+A_2t+A_3=0$, but  on this branch it is
impossible to satisfy $Q=\partial Q/\partial x=0$ if the $e_i$
are distinct.  If the $e_i$ are not distinct, the singularity that
we get when $t^3-3xt+A_2t+A_3=0$ does not affect the part of
the Jacobian of $C$ that is  actually relevant to the $N=2$ model -
it affects the part that comes from $E$.}
\eqn\iin{3t^2-3x+A_2 = 0.}
This can be used to eliminate $x$ from \threecurve, so we reduce
to an equation $H(t)=0$ for some polynomial $H$:
\eqn\hiin{\eqalign{
H=&(A_2-\sum_ie_i)t^4\cr &+A_3t^3+\left({1\over 3}A_2^2-{2\over 3}A_2
\sum_ie_i +\sum_{i<j}e_ie_j\right)t^2 -{1\over 4}A_3^2
+{1\over 27}\prod_{i=1}^3(A_2-3e_i).\cr}}
Singularities of the curve $C$ show up as solutions of
$H=dH/dt=0$, but the converse is not quite true; the equations
$H=dH/dt=0 $ can have solutions at $t=0$
that do not come from singularities
of $C$.  The singular locus of $C$ can be found finally
by computing the discriminant of $H(t)$ and throwing away
a factor that comes from solutions of $H=dH/dt = t = 0$.
One is left with a very complicated polynomial $\Delta(A_2,A_3)$
which we will call the discriminant.

However, for weak coupling a simplification appears.
One can go to weak coupling by setting  $e_i=(0,0,1)$, and then
$\Delta$  turns out to  factor as
\eqn\lnln{\left(4A_2^3+27 A_3^2\right)^2\left(
4(A_2-1)(A_2-4)^2+27A_3^2\right) .}
This coincides with $D_V^2D_H$ if we identify
\eqn\plnln{A_k=(2i)^kW_k.}
The  factor of $(2i)^k$ could be absorbed in rescaling
the Higgs field.
The fact that    $D_V$ is squared but $D_H$ is raised to the first
power presumably has something to do with the fact that extra massless
vector multiplets come in pairs, but extra massless hypermultiplets
come one at a time.  Note that the constants in \plnln\ are fixed
by recovering the formula for $D_V$, and then one automatically
gets $D_H$.

\subsec{Confinement And Higgs Mechanism For $N=1$}

Now we want to discuss a key point of physics: how one sees
confinement and the Higgs mechanism in this formalism.

In \paperi, the elliptic curve controlling the pure $N=2$ theory
for $SU(2)$ developed a node or ordinary double point at certain
points in the complex $u$ plane.  Under a certain mass perturbation
that breaks $N=2$ to $N=1$ supersymmetry, $u$ is locked near one
of these points; monopole condensation occurs, giving a mass gap
to the $N=2$ theory and triggering confinement.

For groups of rank $r>1$, the theory is described
in terms of  a Riemann surface of higher
genus.\foot{Or in general, an abelian variety of higher rank.
We will use the Riemann surface language for simplicity and because
it is adequate for the present case.}  Every time a node develops,
the genus of the Riemann surface drops by one and there appears
a massless monopole  that is charged under one of the $U(1)$'s in the low
energy gauge group.  The massive vacua that will appear in the presence
of the  mass perturbation to $N=1$ -- massive because
of confinement or a Higgs mechanism -- correspond to points in
moduli space at which there are $r$ different nodes, so that
the low energy gauge group is completely broken by monopole
condensation.

In our problem, we start with a genus one surface $E$, and then
determine the physics by an $n$-fold cover $C\to E$, $C$ being
a Riemann surface of genus $n$.  The low energy gauge group has
rank $r=n-1$ (that being the rank of $SU(n)$).  To get a totally
massive    vacuum, $C$ must develop $r$ nodes.  This will reduce
the genus of $C$ to one.
The cover $C\to E$ is still an $n$-fold cover, unramified as $C$ and
$E$ both have genus one.

Totally massive vacua are thus associated with $n$-fold unramified
covers of $E$; moreover, given such a cover, there is precisely
one point in the moduli space of the Coulomb branch at which it
occurs.
To verify the last assertion,
start with an $n$-sheeted cover
$\pi:\tilde{E}\longrightarrow E.$
The claim is that there is a unique $C$ of arithmetic genus $n$ with
normalization
(=desingularization)  $\tilde{E}$, which occurs as a spectral
curve in our
system. First note that the automorphisms of $\tilde{E}$ over $E$
permute the
$n$ points in $\pi^{-1}(p)$ transitively, so it does not matter
which of these we label as the point $q_0$ where the residue is $-(n-1)$.
Having made that
choice, Riemann-Roch guarantees the existence of a unique
function $f$ on $\tilde{E}$ with first order poles at
$\left\{ q_0, q_1, \dots, q_{n-1} \right\}=\pi^{-1}(p),$
and
respective residues $-(n-1),+1, \dots, +1$.
The spectral curve $C$ is then
essentially
the image of $\tilde{E}$ in $E \times \bf{P}^1$under $(\pi,f)$.

Thus to classify the totally massive vacua is the same as classifying
the $n$-fold unramified covers $C\to E$.
This is easily done.  One can realize $E$ as ${\bf C}/\Gamma$
with
$\Gamma\cong {\bf Z}\oplus {\bf Z}$ a lattice
in the complex plane ${\bf C}$.  In that
realization, the $n$-fold cover $C$ is ${\bf C}/\Gamma'$, where
$\Gamma'$ is a sublattice of $\Gamma$ of index $n$.  Any such
sublattice contains $n\Gamma$, which is a sublattice of $\Gamma$
of index $n^2$.  The quotient $\Gamma/n\Gamma$ is an abelian
group isomorphic to $F= {\bf Z}_n\times {\bf Z}_n$.
The quotient $\Gamma'/n\Gamma$ is a subgroup $F'$ of $F$ of index
$n$.  Conversely, every index $n$ subgroup $F'$ of $F$ determines
by this construction a $C$.

So the set ${\cal S}$
of massive vacua (in the presence of the perturbation
to $N=1$) is the same as the set of index $n$ subgroups $F'$ of
$F={\bf Z}_n\times {\bf Z}_n$.
The action of $SL(2,{\bf Z})$ on ${\cal S}$ comes from its action
on $\Gamma$ and thus from the natural action of $SL(2,{\bf Z})$
on ${\bf Z}_n\times {\bf Z}_n$.

In fact, according to `t Hooft's abstract classification
\thooft\ of phases of $SU(n)$ gauge theory,
 the possible massive phases are classified by index
$n$ subgroups of ${\bf Z}_n\times {\bf Z}_n$.
(This is perhaps not as well known as the principles of
the classification of phases and will be explained
in section four.)
Thus, the result in the last paragraph strongly indicates that
this theory (perturbed to $N=1$) realizes every possible massive phase
of an $SU(n)$ gauge theory precisely once.  We will argue this
directly in section four (where we will also discuss some
Coulomb phases of this theory).
The relation to 't Hooft's classification also makes it clear that
the two factors of ${\bf Z}_n$ in $F={\bf Z}_n\times {\bf Z}_n$ can be
considered as labeling magnetic and electric charge; then
the action of $SL(2,{\bf Z})$ on ${\cal S}$ found in the last
paragraph is the expected action of $S$-duality on the magnetic
and electric charges in the vacuum condensate.

For $n=2$, $F$ is equivalent
to  the additive group ${\bf Z}_2\times {\bf Z}_2$
of  points of order two in $E$, and
the choice of an index two subgroup $F'$ is just the choice of
a non-zero point of order two.  Thus our description of the
massive vacua and the $SL(2,{\bf Z})$ action on them
agrees for $n=2$ with the description in \paperii, where these
vacua were identified with the non-zero points of order two.

\bigskip\noindent
{\it Details For $n=3$}

To make this more concrete, we will
 briefly explain how to explicitly find for $n=3$ the
massive vacua, that is the points in moduli
space at which $C$ has $n-1=2$ nodes.  There are three index three
subgroups of ${\bf Z}_3\times {\bf Z}_3$ generated by $(1,x)$ for
$x=1,2,$ or $3$, and one generated by $(0,1)$.  In all, then,
there should be four values of $A_2 $ and $A_3$ at which $C$ has
two nodes.  Let us find them.

 First of all, these points
are all at $A_3=0$.  (More generally, we will see in section four
that the massive vacua are associated with certain $SU(2)$ generators $J$.
As $\Tr\,J^{2r+1}=0$ for any $SU(2)$ generator, the massive
vacua all have $A_{2r+1}=0$ for any $r$ and $n$.)  This being so,
equation \hiin\ simplifies to
\eqn\hhiin{H=(A_2-\sum_ie_i)w^2+\left({1\over 3}A_2^2-{2\over 3}A_2
\sum_ie_i +\sum_{i<j}e_ie_j\right)w
+{1\over 27}\prod_{i=1}^3(A_2-3e_i)}
with
\eqn\jiin{w=t^2.}  We recall that singularities of $C$ correspond
to solutions of $H=dH/dt=0$ with $t\not= 0$.
If we write \hhiin\ as $\alpha w^2+\beta w+\gamma$, then
$\alpha,$ $\beta$, and $\gamma$ are
respectively linear, quadratic, and cubic in $A_2$, then the discriminant
$\beta^2-4\alpha\gamma$ is quartic.  When the $e_i$ are distinct,
this quartic function vanishes at four distinct values of $A_2$.
For such a value of $A_2$, the quadratic form in \hhiin\ vanishes
at a unique value of $w$; the two desired nodes of $C$ are then
at $t=\pm \sqrt w$, with
 $x$ and $y$  determined by \iin\ and \forthree.  Note that this
argument actually determines all the singularities at $A_3=0$;
they are the expected nodes, and nothing else.

\subsec{Cusps}

A cusp is simply a singularity of a Riemann surface that looks
like the singularity of the curve $y^2=x^3$ at $x=y=0$.
A cusp in the Riemann surface governing an $N=2$ supersymmetric system
corresponds apparently to a novel and mysterious
kind of critical point, as discussed in \douglas.

Focussing on the case of gauge group $SU(3)$,
let us count, in the weak coupling regime, how many cusps can be
predicted in our theory because of known physical mechanisms.
First of all, there is a  region, discussed in section        3.1,
in   which the theory flows to the pure $SU(3)$ model, without
hypermultiplets.  This theory has \douglas\ cusps at two
values of $A_2,$ $A_3$.  There is also a second regime that one should
consider.  For this    we recall the   weak coupling discriminant
from     section 3.3.  The $\lambda=0$ discriminant has a factor $D_V$
that vanishes precisely when the theory has an unbroken  $SU(2)$ gauge
symmetry.  It has a factor $D_H$ that vanishes precisely when there
is a massless hypermultiplet.  One finds that $D_V=D_H=0$ at the
two points $A_2=4/3$, $A_3=\pm 16i/27$.  Scaling $\lambda$ to zero in a
neighborhood of those points, one expects to reduce to the $SU(2)$
theory with a massless hypermultiplet in the two-dimensional
representation.  $A_2$ and $A_3$ correspond to the order parameter
$u$ and the bare mass in this theory.
This model was analyzed in \paperii\ and has cusps at three
values of the parameters \ref\aretal{P. C. Argyres, M. R. Plesser,
N. Seiberg, and E. Witten, ``New $N=2$ Superconformal Field Theories
In Four Dimensions,'' to appear.}.  Therefore,
the model under study  here for weak coupling (and therefore
generically) should have cusps at at least $2+2\cdot 3=8$ values
of $A_2,$ $A_3$.  It turns out that this is the exact number for
any value of $\lambda$ away from $0,1,\infty$.

In searching for   cusps at given $\lambda$, it helps to know that
the unfolding $y^2=x^3+\alpha  x+\beta$ has the property that the
discriminant, as a function of $\alpha$ and $\beta$, has a cusp
at $\alpha=\beta=0$.  Conversely, cusps in the     discriminant
reflect cusps in the curve.

So to find cusps in the spectral curve, one can look for cusps
in the discriminant
curve $\Delta(A_2,A_3)=0$, where the discriminant $\Delta$
was defined in section 3.3.
Cusps in the discriminant curve can be found by Maple. One way
is to first compute the discriminant of $\Delta(A_2,A_3)$ regarded
as a polynomial in $A_3$ with $A_2$ as a parameter, and then
look for multiple roots of the resulting polynomial $g(A_2)$.
It turns out that $g(A_2)=h(A_2)^3$ where $h$ is a quartic polynomial
whose zeroes are the $A_2$ coordinates of the cusps; for each
such zero there are two values of $A_3$ (related by the symmetry
$A_3\to -A_3$ ) at which the discriminant curve has a cusp.  If we set
$a=6A_2$, then up to a constant factor the polynomial $h(a) $ is
\eqn\unbu{\eqalign{
h(a) =& a^4-24(1+\lambda)a^3+(192+456\lambda+192\lambda^2)a^2
-\left(512(1+\lambda^3) +2688\lambda
(1+\lambda)\right)a\cr &+(4608\lambda+8784\lambda^2+4608\lambda^3).\cr}}
As $\lambda\to 0$, the four roots of this equation behave as follows.
One has $a\to 0$, corresponding to the two cusps of the pure
$SU(3)$ theory, and the other three have $a\to 8$, the value of
$a$ for $D_V=D_H=0$ where one sees the $SU(2)$ theory with a
massless doublet hypermultiplet.\foot{Incidentally, beyond locating   the
cusps, this computation can be extended to
give  a complete classification
of the singularities of the discriminant.  There are four nodes
corresponding to the massive phases that we discussed above,
eight cusps, and a higher singularity, a tacnode, at infinity.  We do
not know the physical meaning of that latter singularity.}

Now we come to an important point.  The discriminant of the polynomial
$h(a)$ can be readily computed and turns out to be a power of
$\lambda(\lambda-1)$.
In particular it never vanishes except at $\lambda=0 $ or $1$,
two points (equivalent under $SL(2,{\bf Z}))$ that correspond to weak
coupling or $\tau=i\infty$.  Thus, except at infinity in moduli
space, the $A_2$ values
of the four cusp pairs are all distinct.  Moreover,
the two cusps of given $A_2$ are related by $A_3\to -A_3$ and
are distinct since the cusps never have $A_3=0$.  (At the end
of section 3.4 we analyzed all singularities at $A_3=0$ without
meeting cusps.) Hence, for any (finite) $\tau$ the eight cusps
are all distinct.

This has the following significance: it means that there
is a well-defined action of $SL(2,{\bf Z})$ on the set of eight cusps.
\foot{Suppose one starts at some point $\tau$ in the upper half
plane with a particular cusp $C$.  Given $g\in SL(2,{\bf Z})$, one
wants to know how $g$ acts on $C$.  $g$ naturally maps $C$
to a cusp $g(C)$ in the theory with a different $\tau$-parameter
$g(\tau)$.  However, by parallel transport in the upper half plane
from $g(\tau)$ back to $\tau$, one can identify $g(C)$ with a cusp
at $\tau$.  Getting
an unambiguous answer this way depends on the fact that the      cusps
never meet; if several cusps were to meet at $\tau=\tau_0$, the
result of parallel transport would be affected by which way one
wraps around $\tau_0$.}  This is analogous to the fact that
the massive phases also never meet for finite $\tau$, so that
there is an $SL(2,{\bf Z})$ action on the set of massive phases,
which was determined above.  The difference is that the massive
phases are more or less understood, and there was a prediction
for how $SL(2,{\bf Z})$ should act on them; the cusps are not well
understood and there is no prediction to compare to.

The result that emerges for the action of $SL(2,{\bf Z})$ on the
cusps is as follows.  Let $u, v$  be integers defined modulo
three and not both zero.  Note that there are eight possible choices
of the pair $u,v$.  The natural two-dimensional representation
of $SL(2,{\bf Z})$ can be reduced mod three to give an action
by permutation of the eight possible pairs $u,v$,\foot{Think
of $u,v$ as a mod three column vector $\left(\matrix{ u \cr v\cr}\right)$
acted on by an $SL(2,{\bf Z})$ matrix.}
 and we claim that
this is the representation by which $SL(2,{\bf Z})$ acts on the cusps.
Note that this is equivalent to saying that the cusps correspond
to points of order three on $E$.  Also, note that if we divide
by the center of $SL(2,{\bf Z})$ and identify
$u,v$ with $-u,-v$, we would have four equivalence classes corresponding
to the four {\it subgroups} of order three; these are from the above
analysis in one-to-one correspondence with the massive phases.

Though we do not know a natural proof that the $SL(2,{\bf Z})$ action
on the cusps is as stated above, it is possible to verify this
by checking generators and relations.
$SL(2,{\bf Z})$ is generated by the two elements
\eqn\sgen{\eqalign{S&=\left(\matrix{ 0 & 1\cr -1 & 0\cr }\right)\cr
                   T& = \left(\matrix{ 1 & 1 \cr 0 & 1 \cr}\right)\cr}}
with relations $S^4=(ST)^3=1$ and $S^2T=TS^2$.  Note that
$S^2$ is the central element $-1$ of $SL(2,{\bf Z})$, which
acts by $A_3\to -A_3$.  We know that $-1$ does not leave
fixed any cusp (since no cusp is at $A_3=0$), so as a permutation
of the eight  cusps $S^2$ is conjugate to $(12)(34)(56)(78)$.
It follows that $S$ is conjugate to $(1234)(5678)$.  As for
$T$, it is the transformation $\theta\to \theta+2\pi$ on the
underlying $\theta$ angle.  Physically, this is expected to leave
invariant the cusps in the pure $SU(3)$ theory while permuting
the three cusps in the $SU(2)$ theory with the hypermultiplet.
This behavior  can also be seen in the monodromy of the four roots
of $h(a)$ around $\lambda
=0$; one is invariant, and three are permuted.
Thus $T$ is conjugate to the permutation $12(345)(678)$.  It
is a straightforward exercise to check that there is only
one action of $SL(2,{\bf Z})$ by permutation of eight objects
with the conjugacy classes of $S$ and $T$ as given, so the action
of $SL(2,{\bf Z})$ on the cusps is as stated in the last paragraph.

Finally, note that the action of $SL(2,{\bf Z})$          on the
eight cusps is transitive.  This in particular proves that the
critical point associated with the cusp in the pure $SU(3)$ theory
is equivalent to the critical point associated with the cusp
in the $SU(2) $ theory with the hypermultiplet.

\subsec{Higgs to Infinity}

As another check of our model,
we want to consider the flow from $SU(n)$ to a subgroup as $\phi$
becomes large.  If $\phi = B+C/s$, where $C$ is a block-diagonal
matrix that breaks $SU(n)$ to a product $H=\prod_iSU(n_i)\times
U(1)^{k-1}$,
where $n_i$ are the sizes of the blocks in $C$ and $k$ is the number
of blocks, then in the limit of $ s\to 0$, we should reduce
to a product of $SU(n_i)$ theories with free $U(1)$ theories.
This is actually one point where it is more transparent to use
the abstract formulation of section 2 rather than the explicit
equations that we have used so far in the present section.

In section 2, we had a vector bundle $V$ over an elliptic curve $E$,
which decomposes as a sum of line bundles
\eqn\lok{V =  \bigoplus_{i=1}^{n}{L_i}}
There was also a differential $\phi$ on $E$, with values in
the adjoint representation, and a prescribed type of pole at
a  distinguished point $p$
in $E$; $\phi$ is often called the Higgs field.
  The decomposition $\phi = B+C/s$ of the four-dimensional
Higgs field simply corresponds to a decomposition
$\phi=B+C/s$ for the two-dimensional Higgs field, in the following sense.
  In the basis in \lok,
$C$ is the same block-diagonal matrix with constant diagonal
elements that appears in the four-dimensional description,
but $B$ is now different -- it is an adjoint-valued differential on $E$
with a particular     sort of pole (whose characteristic polynomial
will nonetheless ultimately be related to the vacuum expectation value
of the physical Higgs field).

The spectral cover is now given by the equation $\det(t-(B+C/s))=0$;
we want to take the limit of the cover as $s$ goes to zero.
After an obvious rescaling of $t$, we have to look at the equation
$\det(t-C-sB)=0$ in the limit of small $s$.  It may appear that
the limit is just the equation $\det(t-C)=0$, and this is so
if the eigenvalues of $C$ are all distinct, in which case the
unbroken group is just $U(1)^{n-1}$.  The more interesting case
is that in which $C$ has some equal eigenvalues, corresponding
to a non-abelian unbroken group.  In that case, the minors of $B$
in the blocks in which $C$ is constant cannot be disregarded as
they lift the degeneracy otherwise present.  Letting $C_i$ be the
value of $C$ in the $i^{th}$ block, and $B_i$
the corresponding
minor of $B$, and letting $t-C_i=s\tilde t$, the equation in the
$i^{th}$ block becomes $\det(\tilde t-B_i)=0$.
Since $B_i$ is not necessarily traceless, this is the spectral
cover for the $U(n_i)$ theory; its Jacobian
 can be decomposed as the product
of the $SU(n_i)$ abelian variety and a factor of $E$ for each $U(1)$.

There is actually one more point to clarify, which is that $B_i$
automatically has a residue
of the right sort.  That is because the residue condition
was that  the residue of $\phi$ was diagonalizable and
equal to 1 plus a matrix of
rank 1, a condition  inherited by any generic minor such as $B_i$.  Note
that the residue of $B_i$ is automatically traceless,
as the trace of $B_i$ is
 an ordinary differential on $E$ with at most a single pole at $p$;
the pole must be absent as the residues of a differential form
on $E$ always sum to zero.

\subsec{The Surface}

In this section we discuss the symplectic form on the integrable
system. First we give a fairly explicit construction of this form and
discuss its linear dependence on parameters in our setup, and
then we place this in a broader context by reviewing some of the
relevant mathematics literature.

The construction has two parts.  The symplectic
form on the cotangent bundle
$T ^*E = E \times\bf C$
induces a two-form on the total space of  the family of spectral curves.
This in turn determines a two-form on the total space of the family of
Jacobians of spectral curves, which is the total space of the integrable
system, as was explained in section 2.4. We begin with this latter step.

Start with any family $C_b,\      b \in B$
of curves, parametrized by some base $B$, and let $\cal C$
be the total space of the family. Let $J_b$ be the Jacobian
of $C_b$,  let $\cal J$ be the total space of the family of
Jacobians, and let ${\cal C}^{(n)}$ be the relative $n$-th
symmetric product of $\cal C$, that is, the total space of
the family over $B$ whose fiber at $b \in B$ is the $n$-th
symmetric product $C^{(n)}$ of $C$. A two-form $\sigma$
on $\cal C$ determines, in
a natural way, a two-form $\tau$ on $\cal J$.  One way to
see this is to note that the Abel-Jacobi map sends $\cal C$,
as well  as the various ${\cal C}^{(n)}$, to ${\cal J}$.
 It is clear how to
use $\sigma$ to build a two-form on ${\cal C}^{(n)}$; but for
$n>>0$, the Abel-Jacobi map
${\cal C}^{(n)} \longrightarrow \cal J$ is a fiber bundle with
projective spaces for fibers, so the form on ${\cal C}^{(n)}$
must be the pullback of one on $\cal J$. (One needs to check
that the result is independent of the
choice of base point for the Abel-Jacobi map.)

If $\cal C$ happens to be given as a family of curves on a
symplectic surface $S$, there is a natural two-form $\sigma$
on $\cal C$ obtained by pulling back the symplectic form via
the projection map $\cal C \longrightarrow B $. This is the case
for Hitchin's system, where the surface is the cotangent bundle
$T^*E$, and the symplectic form on it is $dx \wedge dt / y$.
(Here $x,y$ are the usual functions on $E$, and $t$ is the vertical
coordinate, on $\bf C$.)
In our case things are slightly more complicated, since the
spectral curves are not contained in
$T ^*E = E \times \bf C$ but rather in its compactification
$T  := E \times P^1$. The form $dx \wedge dt / y$  is meromorphic
on $T$, with second order pole along $\{ t= \infty \} = E \times \infty$.
The spectral curves  intersect the polar locus only at the point
$\{ t= \infty ,  x= \infty \}$, but in a rather complicated way; so it is
natural to transform them to another surface $S$
which is birationally equivalent to $T$. Specifically, we take
 $S$ to be the $P^1$-bundle over $E$: $S := {\bf P} (O_E+O_E( \infty ))$.
It can be obtained from $T={\bf P} (O_E+O_E)$ by blowing up the
problematic point  $\{ t= \infty ,  x= \infty \}$,
then blowing down the original fiber over $\{  x= \infty \}$. When
we map the spectral curves to $S$, they completely miss the section
at $\infty$, and so are contained in the affine part of $S$, which is the
total space of the line bundle $O_E( \infty )$. For the region of this
affine surface near the fiber over $x= \infty$, the natural coordinates
arising from the blowup are $u$ and $s:=tu$, where $u$ is the coordinate
at infinity on $E$ used before (so $x=u^{-2}$ and $y=u^{-3}+...$). The
two-form is
$dx \wedge dt/y = (-2u^{-4}/y) du \wedge  ds$,
so it has a first-order pole along $u=0$ (as well as a second order pole
along $s= \infty $, but our spectral curves never meet this latter locus).
Nevertheless, we claim that the pullback $\sigma$ of this two-form is
everywhere holomorphic, so we can apply the previous construction to
get a symplectic form on the family of Jacobians.  Indeed, consider the
pullback, from a surface $S$ with local coordinates
 $s,u$ to a manifold $\cal C$ mapping to it, of a meromorphic two-form
$du \wedge ds /u$ with first-order poles along $u=0$. The pullback
is holomorphic if  the map is ramified above the polar locus $u=0$.
Back in our situation, the equation of the universal spectral curve,
$$ t^n - {n(n-1)  \over  2}xt^{n-2}+ ... + A_2 t^{n-2}+ ... = 0,$$
when multiplied by $u^n$, becomes
$$ s^n - {n(n-1)  \over  2}s^{n-2} + ... + A_2 s^{n-2} u^2 + ... =0.$$
Choose any of the $n-1$ sheets near $u=0, s=1$ (or the one near
$u=0, s=1-n$). The map which projects this to the $(s,u)$-plane
is indeed ramified above $u=0$,  since the coefficients of each
of the $A_i$ vanish to order $\geq 2$. We conclude that the
pullback $\sigma$ is holomorphic on $\cal C$, and we get our desired
symplectic form on $\cal J$.

In the above construction, we fixed the mass at $m=1$. Consider
what happens when we allow $m$ to vary. The  surfaces $T$ and
$S$ are fixed, but the family of curves in  $S$ appears to depend on
$m$. This dependence can of course be eliminated by applying to
$S$ the automorphism $s \to ms$. The price we pay is that the two-form
$dx \wedge dt/y$ is now multiplied by $m$. Since the two-forms $\sigma$
and $\tau$ on the respective total spaces of spectral curves and spectral
Jacobians are determined from this by linear operations, we conclude
that $\tau$ depends linearly on $m$ as claimed.

The existence of families of symplectic forms depending on linear
parameters is known to extend in several directions. There is a beautiful
result of Mukai
\ref\Mukai{S. Mukai, Symplectic Structure of
the Moduli  Space of Sheaves on an Abelian or K3 Surface,
Invent. Math. {\bf 77} (1984) 101-116},
which says that the moduli space of simple sheaves on a
symplectic surface itself carries a natural symplectic structure.
Mukai applies his construction to the moduli of sheaves on an
Abelian or K3 surface, but with a small modification it applies also
to Hitchin's system, and with some more significant changes, it
turns out to apply to our system as well.

Mukai's result applies to coherent sheaves on a symplectic surface $S$,
whose support could be the entire surface (e.g. vector bundles on the
surface),  or a finite subscheme (this recovers Beauville's construction
of a
symplectic structure on an appropriate resolution of singularities of
the symmetric product of a K3 surface), or as in our case, a curve on the
surface. The result is quite intuitive in the case of finite support: the
symplectic form on $S$ induces one on the product $S^n$, which clearly
descends to a symplectic form on a dense open subset of the symmetric
product; the only remaining issue is to check that this extends
symplectically (i.e. is nowhere degenerate)
on a particular compactification of this open subset, which turns
out to be the Hilbert scheme parametrizing appropriate sheaves
on $S$. In  the general case, Mukai identifies the tangent space to moduli
at a simple sheaf $F$  with the vector space
$$Ext^1_{O_S}(F,F),$$
and notes that any two-form $\sigma \in
H^0(\omega_S)$ determines an alternating bilinear map:
$$
Ext^1_{O_S}(F,F) \times Ext^1_{O_S}(F,F) \to
Ext^2_{O_S}(F,F) \rightarrow H^2(O_S)
\rightarrow  H^2(\omega_S) \approx \bf C,
$$
hence a two-form on moduli, which is non-degenerate for general duality
reasons.
In general, one still needs to check that this form is closed; when
the
moduli space is known to be smooth and projective, this follows for free
from the Kahler package.

Hitchin's system fits as the special case where the surface  is the
cotangent bundle $T^*E$ of a curve $E$, and the sheaves
are (the extension to the surface of) line bundles on (spectral) curves .
In our case, the sheaves are supported on curves contained in the
surface $S$ which is the total space of the non-trivial line bundle
$O_E(p)$,
and this surface is {\it not} symplectic. Rather, it is Poisson,
i.e. it has a natural two-vector field (the inverse of
the meromorphic two-form
$dx \wedge dt/y = (-2u^{-4}/y) du \wedge  ds$) which is non-degenerate
only generically, away from $u=0$. Now a modification of Mukai's argument
(observed  by Tyurin, Mukai, and Markman) shows that
the moduli of sheaves on a Poisson
surface itself carries a Poisson structure. The space
we are interested in then appears as one symplectic leaf of this general
Poisson space
(or as a one-parameter family of such leaves, if
we allow the mass to vary rather than fixing it at $1$).

\newsec{Analysis Of The Massive Phases}

This final section is devoted mainly to explaining some points of physics
that are needed to properly understand the analysis of the
totally massive vacua given in section three.  We will also
make a few remarks on Coulomb phases.

We first make a few simple observations on 't Hooft's abstract
classification of the possible vacua.  Then we explain from a weak
coupling point of view why the particular theory under study
has the beautiful property of realizing each of these vacua precisely
once.  Then we turn to the Coulomb phases.

\subsec{Classification Of Phases}

Consider an $SU(n)$ gauge theory in which all fields
transform trivially under the center of $SU(n)$, which is
isomorphic to ${\bf Z}_n$.  An example is the perturbed $N=4$
theory discussed in this paper.  One might say that in such
a theory the gauge group could really be taken as $SU(n)/{\bf Z}_n$.

The group of possible external charges by which this theory
might be probed, modulo the charges of the dynamical fields,
is the group of characters of ${\bf Z}_n$
 and is isomorphic  to ${\bf Z}_n$.  We refer
to this ${\bf Z}_n$ as the group of electric charges.

In addition, if one is in four dimensions, then as the fundamental
group of $SU(n)/{\bf Z}_n$ is    again ${\bf Z}_n$, the possible
$SU(n)/{\bf Z}_n$ bundles on a two-sphere at spatial infinity
correspond to elements of
 ${\bf Z}_n$.  This ${\bf Z}_n$  classifies the possible
magnetic charges by which the theory can be probed.
The group $F={\bf Z}_n\times {\bf Z}_n$ of these possible
magnetic and electric charges (we  will write the magnetic charge
first and the electric charge second) plays a fundamental role
in 't  Hooft's abstract classification of the possible phases
of an $SU(n)$ gauge theory.

For every  point $(a,b)\in F$, and every loop $C$ in space-time,
one can define a loop operator $W_{a,b}(C)$  ($W_{(0,1)}(C)$ is the
Wilson loop operator, and $W_{(1,0)}(C)$ is often called the
't Hooft loop operator; $W_{a,b}(C)$ is essentially the product
of $a$ copies of the 't Hooft operator and
$b$ copies of the Wilson operator).
In computing the algebra of the $W$'s,
one meets a certain natural skew form $\langle~,~\rangle$
on $F$, with values in ${\bf Z}_n$.  $\langle ~ , ~\rangle$ can be
defined by saying that  for $x=(a,b)$ and $y=(c,d)$,
\eqn\jsnn{\langle x,y\rangle =ad-bc ~{\rm modulo}~n.}

The importance of this skew form in the theory begins with
 the following.
If a field of charge $x$ condenses in the vacuum, then (as 't Hooft
argues) any field $y$ with $\langle x,y\rangle \not=0$ is confined.
Therefore, confinement of electric charge follows from
condensation of magnetic charge, and vice-versa.  Of course,
this generalizes the usual Meissner effect in superconductors.

't Hooft further proves that if charges $x$ and $y$  both condense,
then $\langle x,y\rangle = 0$.  It follows immediately that if
$F'$ is the group generated by the condensed charges, then
the order $d$ of $F'$ is at most $n$, and in fact  $d$ is a divisor
of $n$.  (Conversely, every $F'$ whose order is a divisor of $n$
automatically has the property that $\langle x,y\rangle = 0 $ for
$x,y\in F'$.  For $F'$ of order $n$ this follows from the classification
of the index $n$ subgroups below; the general case is similar.)

In this situation, let $F''$ be    the group consisting of all $y$
such that $\langle x,y\rangle = 0$ for all $x\in F'$.  Then
$F''$ contains $F'$, and $F''$ is equal to $F'$ if and only if
the order of $F'$ is $n$.  In fact, the order of $F''$ is $n^2/d$,
and the index of $F'$ in $F''$ is $(n/d)^2$.  Let $L=F''/F'$.
Since the unconfined charges are those in $F''$, and the charges
in $F'$ have condensed in the vacuum, $L$ classifies the unconfined
external charges by which the low energy theory can be probed.

For the $L$ equivalence classes of low energy
charges,  one can still use  the 't Hooft-Wilson loop operators
 $W_{a,b}(C)$.
They obey the same algebra as before,  still governed
by the pairing in \jsnn, except that since we are only measuring
the charges modulo $F'$, the pairing is now well-defined only
modulo $n/d$.

't Hooft, however, proves that as long as there are such non-trivial
pairings observed in the low energy theory, there cannot be a mass gap;
one necessarily has a ``Coulomb phase.''
The massive vacua (that is, the vacua with a mass gap) therefore
have $d=n$;  they are classified by the order $n$ subgroups
$F'$ of $F={\bf Z}_n\times {\bf Z}_n$.
Any operation that one would want to call electric-magnetic duality
should be a  subgroup of $SL(2,{\bf Z})$ that
permutes the massive vacua through the natural action
of $SL(2,{\bf Z})$ on ${\bf Z}_n\times {\bf Z}_n$.

This is precisely the structure that we found in section three:
the massive vacua of our theory are classified by index $n$ subgroups
of ${\bf Z}_n\times {\bf Z}_n$, permuted in the natural way by
$SL(2,{\bf Z})$.  This strongly suggests that each and every massive
phase allowed abstractly is realized precisely once in the theory
that we are studying.  We will give further evidence for that
interpretation below.

Before doing so, however, let us make a few simple observations
on the index $n$ subgroups of ${\bf Z}_n\times {\bf Z}_n$.  For
$n$ prime, there are precisely $n+1$ of these, generated by
$(1,x)$ for some $x\in {\bf Z}_n$ or by $(0,1)$.  (In the present
context, condensation of $(1,x)$ corresponds to confinement or
oblique confinement, while condensation of $(0,1)$ is the Higgs
mechanism.)  If $n$ is not prime, the structure is more complicated.
For every positive divisor $d$ of $n$, and every $b$  with
$0\leq b\leq d-1$, there is an  index $n$
subgroup of ${\bf Z}_n\times {\bf Z}_n$ generated by
$x=(n/d,b)$ and $y=(0,d)$.
It is easy to show that all the index $n$
subgroups are of this form for some $d $ and $b$,\foot{
Let $F'$ be a subgroup of ${\bf Z}_n\times {\bf Z}_n$ of order
$n$.  Let $H$ be the subgroup of ${\bf Z}_n$  consisting of
all $a $    such that $(a,b)\in F'$ for some $b$.  If $H$ is of order
$d$, then $d$ is a divisor of $n$ and $H$ is generated by $n/d\in
{\bf Z}_n$.  If so, then $x=(n/d,b)\in F'$ for some $b$.   For
$F'$ to have order $n$, the kernel of the homomorphism
$F'\to H$
must have order $n/d$, and this means that $F'$ must
contain $y=(0,d)$.  We can therefore assume that $F'$ contains
$x=(n/d,b)$ with some $b\leq d-1$, as desired.  Note that $d$ can
be described as the smallest integer such that $(0,d)\in F'$.}
so the number of possible massive phases is the sum of the positive
divisors of $n$.

\subsec{Weak Coupling Analysis}

Now we turn to the weak coupling analysis of our particular
theory -- the $N=2$ theory with a massive matter hypermultiplet.
{}From  an $N=1$ point of view, this theory contains an adjoint
superfield $\phi$ related by $N=2$ to the gauge field, and two
additional adjoint superfields $B,C$ in the hypermultiplet.
The superpotential of the $N=2$ theory (including the bare
mass term for the hypermultiplet) is
\eqn\jnsln{W=\Tr (\phi[B,C]+  BC).}
An $N=1$  theory that has massive phases is obtained if one  perturbs
this by an arbitrarily small mass term for $\phi$.  The
superpotential is then
\eqn\ojnsln{W=\Tr (\phi[B,C]+  BC+\epsilon\phi^2)}
for some complex $\epsilon\not= 0$.

To find the classical vacua, one must find the critical points
of $W$ modulo $SL(n,{\bf C})$, a task which is invariant under
the action of $GL(3,{\bf C})$ on the three objects $\phi,B,C$
and under rescaling of $W$ by a constant.  Via such operations,
one can convert the superpotential into
\eqn\ooks{W=\Tr\left({1\over 2}(X^2+Y^2+Z^2)-X[Y,Z]\right).}
The equations for a critical point are now
\eqn\unn{\eqalign{[X,Y]& = Z \cr
                  [Y,Z]& = X \cr
                  [Z,X]& = Y .\cr}}
As  was noted
in essentially the present context in
\vafa, these equations
are the commutation relations of the Lie algebra of $SU(2)$.
Thus,  after dividing by $SL(n,{\bf C})$, the vacua are parametrized
by the isomorphism classes of homomorphisms $\rho$ from $SU(2)$
to $SU(n)$.  Since such homomorphisms have no infinitesimal deformations,
in any such vacuum, the chiral superfields all have bare masses;
any massless particles come from the gauge multiplets.

Given $X,Y,$ and $Z$ determining
 a homomorphism $\rho:SU(2)\to SU(n)$, the $SU(n)$ gauge group
is spontaneously broken at the classical level to the subgroup
$H$ that commutes with the image of $\rho$.
Since the chiral superfields are massive, the low energy theory
is the pure $N=1$ gauge theory with gauge group $H$.
If $ H$ is semi-simple,
confinement will occur at low energies, and one will get a totally
massive phase.  On the other hand, if $H$ contains $U(1)$ factors,
then one will have a Coulomb phase (for these $U(1)$'s) at low
energies, and there will be no mass gap.

$\rho$ defines an $n$-dimensional representation of $SU(2)$
(possibly trivial) which can be decomposed as a sum of irreducible
pieces.  If there are {\it distinct} irreducible pieces in this
representation, then $H$ will contain $U(1)$ factors and there will be
no mass gap.  A mass gap will occur precisely when $\rho$ is the
direct sum of $d$ copies of the $n/d$-dimensional irreducible
representation of $SU(2)$, and then $H=SU(d)$.  The
$N=1$ $SU(d)$  gauge theory without chiral matter has
$d$ vacua.  Here $d$ can be an arbitrary positive divisor of $n$.  So
-- with every such positive divisor $d$ contributing $d$ vacua --
the number of massive vacua in this theory is the sum of the
positive divisors of $n$.

Since this coincides with the total number of massive phases
possible in $SU(n)$ gauge theory, as counted above, it is natural
to think that what is happening in this theory is that each massive
phase is appearing precisely once.  To justify this, note that in  a
 phase that is Higgsed at the classical level to $H=SU(d)$,
an electric charge $(0,c)$ is unconfined (and so is contained
in $F'$) if and only if it transforms trivially under
the  center of $H$, that
is if and only $c$ is divisible by $d$.
So in view of the classification
of $F'$ given in a footnote above, $F'$ is generated by $(0,d)$
and $(n/d,b)$, with some $0\leq b\leq d-1$.  (Note that the fundamental
magnetic charge of the $SU(d)$ theory corresponds to magnetic charge
$n/d$ in $SU(n)$, because of the way $SU(d)$ is embedded in $SU(n)$.)
As the symmetry $\theta\to\theta+2\pi$ of the low energy $SU(d)$
theory maps $(n/d,b)$ to $(n/d,b+1)$, each value of $b$ appears,
and therefore every abstractly possible massive phase is
concretely realized in our theory.

\subsec{  Coulomb Phases Of The $N=1$ Theory}

We have analyzed in some detail the massive phases that appear
in the theory perturbed  by the $\Tr\,\phi^2$ superpotential,
 and located them
in the exact analysis of section three.  Coulomb phases are, however,
also of interest, and one would like to compare the semiclassical
analysis of Coulomb phases to the exact description.  We will do this
only for $n=3$.  (For $n=2$ there are no Coulomb phases after perturbation
by $\Tr\,\phi^2$.)

For $n=3$, $H$  fails to be semisimple only when $\rho$ is the
direct sum of a two-dimensional representation of $SU(2)$ and a trivial
representation.  There is therefore precisely one vacuum of the
perturbed theory in a Coulomb phase.  Let us try to locate it
in the  exact solution.

In the exact solution, there is a divisor $D$ in moduli space
-- one might call it the quantum discriminant -- on which a massless
charged hypermultiplet appears.  Because of reasoning explained
in \paperi, any vacuum that  survives when the perturbation
$\Tr\, \phi^2$ is turned on is located somewhere on $D$.
As sketched in section 3.5,
singularities of $D$ are either normal crossings (where
the curve $C$ has two nodes) or cusps.  The normal crossings
give the massive vacua, and \argyres\ the cusps
do not survive when the $\Tr\,\phi^2$ perturbation is added.
We are therefore dealing with a vacuum associated with
 a smooth point of $D$.

 $D$ can be described as in section 3.3
 by an  explicit equation $\Delta(A_2,A_3)=0$.
At a smooth point of $ D$, only one monopole hypermultiplet
-- described in an  $N=1$ language by a pair of chiral
superfields $M,\,\tilde M$ -- is relevant in the low energy
description.  The superpotential is thus $W=\Delta(A_2,A_3)M\,\tilde M$.
In the presence of the perturbation, since $\Tr\,\phi^2$ is simply
a multiple of $A_2$, we need to look at the superpotential
\eqn\knln{W=\Delta(A_2,A_3)M\tilde M + \epsilon A_2.}
The equations for a critical point of $W$ are
\eqn\bbn{\eqalign{\Delta(A_2,A_3)M=\Delta(A_2,A_3)\tilde M & = 0\cr
                  {\partial \Delta
                    \over \partial A_2}M\tilde M +\epsilon &
               = 0 \cr
             {\partial \Delta\over \partial A_3}M\tilde M & = 0.\cr}}
The second equation forces  $M\tilde M\not= 0$, so the other
equations imply
\eqn\nubbo{\Delta={\partial \Delta\over \partial A_3}=0.}
Suppose that by adjusting $A_2,A_3$ one finds an isolated,
nondegenerate solution
of those two equations.
Assuming that
$\partial \Delta /\partial A_2\not= 0$ (otherwise one is at
a singularity of the discriminant, a case that we have excluded),
the  second equation in \bbn\ then
 determines the gauge-invariant product $M\tilde M$, which vanishes
as $\epsilon\to 0$.
  After setting to zero
the $D$ terms of the low energy theory and dividing by the gauge
group, one gets precisely one vacuum from each isolated
solution of \nubbo\
with $\partial \Delta/\partial A_2\not= 0$.

There should be for generic coupling
precisely one value of $A_2 $ and $A_3$ obeying these conditions,
since we have seen semiclassically that the theory with $\epsilon\not= 0$
has precisely one Coulomb vacuum.  To exhibit the existence of such a
state, look at the discriminant in the weak coupling limit.
This was found in \lnln\ to be
\eqn\lnlnl{\Delta
=(4A_2^3+27 A_3^2)^2\left(4(A_2-1)(A_2-4)^2+27A_3^2\right).}
We notice that the conditions $\Delta=\partial \Delta/\partial A_3=0$,
$\partial \Delta/\partial A_2\not= 0$ are obeyed precisely
at $A_3=0, $ $A_2=1$.  Moreover, at this value of $A_2 $ and $A_3$,
the functions $\Delta$ and $\partial \Delta/\partial A_3$ vanish only
to first order,
so if $\Delta$ is perturbed, there will be a nearby solution of
these equations at which $\partial \Delta/\partial A_2$ will be still
non-zero.  Thus, we have found a Coulomb vacuum in the exact solution.

\listrefs
\end